\documentclass[preprint]{aastex}
\usepackage{color}
\usepackage[dvipdfm,pdfstartview=FitH,CJKbookmarks=true,
bookmarksnumbered=true,bookmarksopen=true]{hyperref}

\begin{document}

\title{The Oxygen Features in Type Ia Supernovae and the Implications for the Nature of Thermonuclear Explosions}

\author{Xulin Zhao\altaffilmark{1,2}, Keiichi Maeda\altaffilmark{2,3}, Xiaofeng Wang\altaffilmark{1}, Lifan Wang\altaffilmark{4},
Hanna Sai\altaffilmark{1}, Jujia Zhang\altaffilmark{5}, Tianmeng Zhang\altaffilmark{6}, Fang Huang\altaffilmark{1}, and Liming Rui\altaffilmark{1}}

\altaffiltext{1}{Physics Department and Tsinghua Center for Astrophysics, Tsinghua University, Beijing, 100084, China:
wang\_xf@mail.tsinghua.edu.cn; zhaoxl11@mails.tsinghua.edu.cn}
\altaffiltext{2}{Department of Astronomy, Kyoto University, Kitashirakawa-Oiwake-cho, Sakyo-ku, Kyoto 606-8502, Japan: keiichi.maeda@kusastro.kyoto-u.ac.jp}
\altaffiltext{3}{Kavli Institute for the Physics and Mathematics of the Universe (WPI), University of Tokyo, 5-1-5 Kashiwanoha, Kashiwa, Chiba 277-8583, Japan.}
\altaffiltext{4}{Mitchell Institute for Fundamental Physics and Astronomy, Texas A\&M University, College Station, TX 77843, USA}
\altaffiltext{5}{Yunnan Astronomical Observatory of China, Chinese Academy of Sciences, Kunming, 650011, China}
\altaffiltext{6}{National Astronomical Observatory of China, Chinese Academy of Sciences, Beijing, 100012, China}
\begin{abstract}

The absorption feature O~I $\lambda$7773 is an important spectral indicator for type Ia supernovae (SNe~Ia) that can be used to trace the unburned material at outer layers of the exploding white dwarf. In this work, we use a large sample of SNe~Ia to examine this absorption at early phases (i.e., $-$13 days $\lesssim$ t $\lesssim$ $-$7 days), and make comparisons with the absorption features of Si~II $\lambda$6355 and Ca~II near-infrared (NIR) triplet. We show that for a subgroup of spectroscopically normal SNe with normal photospheric velocities (i.e., v$_{si}$ $\lesssim$ 12,500 km s$^{-1}$ at optical maximum), the line strength of high velocity feature (HVF) of O~I is inversely correlated with that of Si~II (or Ca~II), and this feature also shows a negative correlation with the luminosity of SNe Ia. This finding, together with other features we find for the O~I HVF, reveal that for this subgroup of SNe~Ia explosive oxygen burning occurs at the outermost layer of supernova and difference in burning there could lead to the observed diversity, which are in remarkable agreement with the popular delayed-detonation model of Chandrasekhar mass WD.

\end{abstract}


\section{Introduction\label{S_introduction}}

Type Ia supernovae (SNe~Ia) are important in cosmology and astrophysics, not only for the key roles that they played in the discovery of accelerating universe \citep{ri98,pe99}, but also for their contribution to production of heavy elements in the Universe. However, the nature of their progenitor systems is still controversial \citep[e.g.,][ and references therein]{mao14}. It is widely accepted that SNe~Ia are a result of a thermonuclear explosion of carbon oxygen (C/O) white dwarfs (WDs), but there are two major channels proposed so far that could lead to such an explosion. One is an explosion of a WD which accretes hydrogen or helium from its non-degenerate companion \citep[single degenerate scenario,][]{wh73,nom82,nom97}, with supporting evidences from the detections of circumstellar material (CSM) around some SNe~Ia \citep{ham03,ald06,pa07,st11,di12,mag13,sil13}. The other is a merger of two WDs \citep[double degenerate scenario:][]{ib84,we84}, which has recently gained special attention due to the observational findings that there are no companion signature found for some SNe~Ia, i.e., the nearby object SN 2011fe and some supernova remnants such as SN 1006 and SNR 0509-67.5 in LMC, down to the luminosities much fainter than the Sun \citep{li11, blo12, bro12, her12, sch12}.

There are also diversities in spectroscopic and photometric properties of SNe~Ia, and it has been discussed whether all (or spectroscopically normal) SNe Ia are originated in a single evolutionary path or there are multiple populations. For example, there is a so-called high velocity (HV) subclass of SNe Ia \citep{wa08,wa09a}, with redder peak B $-$ V colors and slower late-time decline rates in bluer bands relative to the normal velocity (Normal) subclass\footnote{This spectroscopic classification is based on the photospheric velocity measured from Si II $\lambda$6355 lines in the near-maximum-light spectra\citep{wa09a}.}, and it is interesting to explore whether they represent different populations. The observed differences between the HV and Normal SNe~Ia could be at least partly attributed to be a geometric effect of an asymmetric explosion \citep{mae10, mau10}. However, such observed diversities are also found to be linked to their birthplace environments to some extent \citep{wa13}, suggesting that SNe~Ia may arise from multiple classes of binary evolution. Taking together, it could mean that there are two populations in the HV subclass in which one is related to the Normal SNe Ia while the other is not.

The other long-standing problem is the physical process of SN Ia explosions \citep{hi00}. Theoretically, the thermonuclear reaction disrupting the star may propagate at a range of speeds from subsonic deflagration to supersonic detonation \citep{nom84,kh91,fin10,pak12}. The most popular scenario is the delayed-detonation model \citep{kh91} which assumes that a strong detonation unbinds the whole progenitor WD, after a deflagration which produces stable Fe-peak elements near the center of the WD. Another popular model involves the double-detonation mechanism \citep{liv90,fin07,sim12,she14}. In this model, explosive He burning is induced on the WD surface by compressional heating if the accreted He amount is sufficiently large, and this creates a shock-wave penetrating into the WD core. Once the shock wave is converged near the center, a powerful detonation can be triggered and it will lead to the explosion unbinding the whole WD. Existing observational diagnostics offer some clues to the explosion models, while this topic is still controversial.

The high-velocity absorption features (HVFs) can provide clues to the burning processes in the outer layers of the exploding WD, though their origins are still debated. This is a high-velocity ($>$16,000 km s$^{-1}$) component that is likely formed in regions lying above the photosphere \citep{hat99,ma05a}. The HVFs of Ca~II near-infrared (NIR) triplet and Si~II~6355 \AA\ in SNe Ia have been systematically examined using both early-phase and near-maximum-light spectra \citep{mag12,ch14,sil15,zhao15}. These studies indicate that the abundance distributions of Si and Ca are strongly related, likely as a result of the same burning process responsible to create Si and Ca. Since Si and Ca are produced basically in the same burning layer, their relative strengths are not expected to be sensitive to the typical burning process they experienced. It is thus interesting to make comparisons with absorption features created by elements whose abundance pattern is different from Si and Ca in different burning layers. The C~II $\lambda$6580 could be an important indicator of unburned fuel in the progenitor and the observations of this feature have been discussed in many literatures \citep{tho11, pa11, bl12, fo12, si12c, mag14, hsi15}. However, the C~II $\lambda$6580 absorption is usually very weak and difficult to be quantified in line strength and velocity distribution. On the other hand, the O~I $\lambda$7773 line is relatively strong but not explored so far. We
thus propose using this feature as a tracer of unburned materials in this paper, although part of oxygen is also produced via carbon burning.

The presence of HVF in O~I $\lambda$7773 absorption was only reported in the earliest spectra of SN 2011fe in M101 but it was observed to disappear rapidly \citep{nu11}. Detection of such an O-HVF is critically important, since Si and Ca are produced from O via the oxygen burning, and the comparison between the HVF of O and those of Si and Ca can be potentially used to confirm their associations. In the Si burning layer, the oxygen is fully consumed; in the O burning layer, all the O, Si, and Ca can have large abundances; in the C burning layer, the oxygen is abundant but the Si and Ca are under-abundant; and finally in the unburned layer again, the ratio of O to Si or Ca is large. Therefore, we may obtain information about the typical burning process encountered by the layer producing the HVFs, and hence the still-unclarified origin of the outermost materials in SN Ia.

This paper is organized as follows. In \S\ref{S_measure}, our data samples and the methods for measuring the spectral parameters are introduced. In \S\ref{S_correlation}, correlations between spectroscopic parameters of oxygen (O~I $\lambda$7773) and those of other elements (Si~II $\lambda$6355, Ca~II~NIR triplet, and C~II $\lambda$6580) are investigated. Correlations between spectroscopic features of O~I $\lambda$7773, Si~II $\lambda$6355, Ca~II~NIR triplet and luminosity indicator $\Delta$m$_{15}(B)$ are also examined and analyzed in this section. Origins of the high velocity features and the constraints on the explosion models are discussed in \S 4. The paper is closed in \S\ref{S_conclusion}.

\section{Dataset and Methods of Measurement\label{S_measure}}

\subsection{Sample}\label{S_sample}
Since the HVFs of SNe Ia are only prominent in their early spectra, we choose the sample with spectroscopic observations at t $\lesssim$ $-$ 7 days from the B-band maximum light. From the published spectral database from the CfA supernova program \citep{mat08,bl12}, the Berkeley supernova program \citep{si12a}, Carnegie Supernova Project \citep[CSP:][]{fola13}, and the unpublished dataset from the Tsinghua supernova program\footnote{This dataset contains nearly 1000 spectra for over 200 SNe (Rui et al. in preparation), obtained with the Lijiang 2.4-m telescope (+YFOSC) of Yunnan Astronomical Observatories and the Xinglong 2.16-m telescope (+BFOSC) of National Astronomical Observatories of China.}. The telluric absorption features were removed from almost all of the above spectral sample using the standard-star spectra obtained with the same slit used for the SN observations. Nevertheless, this procedure often leaves residuals from the strongest telluric bands near 7620\AA. With the multi-gaussian fit addressed in next subsection, we examine the blending of the residual telluric absorption with the O~I $\lambda$7773 absorption in the spectra of our SN Ia sample. Those spectra showing larger degrees of blending between these two absorptions (see \S 2.2) are excluded in our analysis. Moreover, the near-maximum-light Si~II $\lambda$6355 velocity is also required for further classification of our sample. This leads to a total of 143 early-time spectra (with t $\lesssim$ $-$7 days from the B-band maximum light) that cover the wavelength region of O~I $\lambda$7773 absorption. These spectra belong to 62 SNe Ia, including 55 spectroscopically normal SNe Ia, 2 SN 1991T-like \citep{fil92, phi92} and 5 SN 1999aa-like SN Ia \citep{li01} (hereafter these two subclasses are abbreviated to 91T/99aa-like SNe Ia). The spectroscopically normal objects can be further divided into 42 Normal SNe~Ia and 13 HV ones.

The above sample of SNe~Ia are mostly included in our previous study \citep{zhao15}, where their photometric and host-galaxy parameters (i.e.,$\Delta$m$_{15}$($B$), $B_{max}-V_{max}$, spectroscopic subclassification, host-galaxy type, $K$-band magnitude of the host galaxy etc.) can be found. The sources of the light curves are from the Harvard CfA SN group \citep{hi09,hi12}, the CSP \citep{con10,str11}, the Lick Observatory Supernova Search \citep[LOSS:][]{ga10}, and our own database.

\subsection{Measurement Procedure}\label{S_measurement}
The measurement procedure of O~I absorption is overall similar to that applied to measuring the Si~II~6355/5972 and Ca~II near infrared     (NIR) triplet absorptions \citep{zhao15}, but with more gaussian components being used to fit more absorptions. The absorptions identified around O~I $\lambda$7773 include three O I components (two high-velocity features and one photospheric component), three residual telluric components (i.e., at $\lambda$s 7186, 7606, 7630\AA), and a possible Mg~II $\lambda$7890 (with a velocity at 10,000 km s$^{-1}$ $\sim$ 14,000 km s$^{-1}$). These absorption components can be represented using the following multi-component Gaussian function:
\begin{equation}\label{1}
f(\lambda)=\sum_{i=1}^6 A_i
exp(-\frac{(\lambda_i-\lambda_i^{0})^2}{2\sigma_i^2})
\end{equation}
where subscript $i$=1$-$6 denote the PHO component of O~I $\lambda$7773, the "main" HVF (hereafter HVF-I), the possible "second" HVF (hereafter HVF-II), and the telluric absorptions at 7168\AA, 7605\AA\ and 7630\AA, respectively. In the above equation, $A_{i}$ is the amplitude (i.e., the maximum absorption), $\sigma_{i}$ is the dispersion (i.e., the standard deviation in a normal distribution), and $\lambda_i^{0}$ is the laboratory wavelength of specific spectral line. Note that the positions of the telluric absorptions (and thereby the $\lambda_{i}^{0}$) are blueshifted relative to the rest-frame wavelengths due to that the spectrum was corrected for the redshift of the host galaxy.

In the fitting, the pseudo-continuum is defined as a straight line connecting the line wings on both sides of the absorption, which is determined through repeated visual inspections and careful manual adjustments in order to better counter the noise. As usual, the line strength of an absorption feature is quantified by the pseudo-equivalent width (pEW). From equation \ref{1}, we have:
\begin{equation}\label{2}
pEW=\sum \sqrt{2\pi} A_i \sigma_i
\end{equation}
where the summation is performed for appropriate components.

Before fitting, we smoothed the observed spectrum around the O~I $\lambda$7773 absorption (i.e., covering the wavelength range from 7000 to 7800\AA) with a locally weighted scatter-plot smoothing method \citep{cle79} in order to reduce the effect of noise spikes on the fitting results, as the O~I $\lambda$7773 line (especially the high-velocity feature) is usually weak in early-time spectra of SNe Ia.
In our previous work \citep{zhao15}, the PHO velocity Si~II~5972 was used as an initial condition to constrain that of Si~II $\lambda$6355 because the HVF and PHO component of the latter feature are often seriously blended. In determining the velocity of O~I $\lambda$7773 line, however, we did not use the velocity condition from Si~II $\lambda$5972 or Ca~II NIR triplet lines because the HVF feature and PHO component of O~I $\lambda$7773 line are distinctly separated (i.e., by about 6000 km s$^{-1}$) and their locations can be better determined in early time spectra, as shown in Figure 1. The only exceptions are for several HV SNe Ia where their HVF and PHO components show {\bf more severe blending} (i.e., SN 2002dj) and the velocity of Si~II $\lambda$5972 is used to help better locate the PHO component of O~I $\lambda$7773.

As most of the published spectral data do not have the accompanied flux errors, we adopt the $R$-squared statistic to fit the parameters of absorption features such as the amplitude $A_{i}$, the dispersion $\sigma_{i}$, and velocity of absorption minima $v_{i}^{0}$. This fitting procedure is done through the Curve Fitting Toolbox in Matalab, where regression analysis can be conducted using the library of various models provided. Generally speaking, the fittings were well-performed and objectively determined by the absorption profiles, with the typical $R$-squared value being 0.97 for our sample. There are some reasons that the multiple gaussian fitting applied in our analysis can be well done without over/under-fitting. Firstly, the blending between HVFs and PHO components of O I $\lambda$7773 is much weaker than that seen in Si II $\lambda$ 6355 or Ca II NIR triplet. Secondly, the contaminations from residual telluric absorptions are not very serious for our selected sample. For example, the blending ratio between the telluric absorption and the photospheric component of O~I $\lambda$7773 is overall small for our sample (see discussions below). Thirdly, in most cases, the possible Mg II $\lambda$7890 is not strong (with pEW $\lesssim$ 10\AA) and it is well separated from the PHO component of O I $\lambda$7773.

Figure 1 shows the absorption features centering at O~I $\lambda$7773 in the t$\approx$$-$10 days spectra of some representative SNe Ia.  Demonstration of applying multiple-gaussian fit is shown in the plot, where the photospheric (PHO) absorption of O~I $\lambda$7773 (at $\sim$ 12,000 km s$^{-1}$) is accompanied by two additional absorptions on the blue side. These two absorptions can be attributed to the HVF-I and HVF-II of oxygen at velocities $\sim$18,000 km s$^{-1}$ and $\sim$22,000 km s$^{-1}$, respectively, and this identification is secured by comparisons with the SYNOW fit as shown in Figure 2 \citep{fish95}. We notice that different absorption components of O~I line are clearly separated from each other for Normal SNe Ia, while the O-HVF absorptions tend to blend with the PHO component in some HV SNe Ia such as SN 2002bo and SN 2002dj.

In this work, errors of spectral fit were retrieved from the Matlab function "confint". This function calculates the confidence bounds of the fitting results using t$\sqrt{((X^{T}X)^{-1}s^{2})}$, where $X$ is the Jacobian of fitted values, $X^{T}$ is the transpose of $X$, $s^{2}$ is the mean squared error, and $t$ is calculated using the inverse of Student's cumulative distribution function. The errors of A$_{i}$ and $\sigma_{i}$ are then used to compute that of the pEW using Eq.(2), and the error of $\lambda_{i}$ is converted to that of the line velocity. Since the fitting is applied to the smoothed spectrum, the noise of original spectrum is further factored into the above errors by taking residual standard deviation between the original and smoothed data roughly as 1-$\sigma$ error in spectral fitting. The typical error from spectral noise is about 3\% for our sample, and the corresponding error in pEW is about 1.0\AA. Additional velocity error is also considered for the fits to the smoothed spectra, which is taken to be velocity shift from gaussian fits to the smoothed and original spectra \footnote{The fit to the original spectra can be done for 87 of 143 for our sample. The resultant
velocities (and pEW) are well consistent with those from fits to the smoothed spectra, with the adjusted R-squared coefficient for linear correlations being 0.91 for the PHO component and 0.83 for the O-HVF component, respectively.} or the standard error of these two fits (i.e., 460 km s$^{-1}$ for the PHO component and 515 km s$^{-1}$ for the HVF for 87 spectra) when the Matlab code fails to fit the original spectra. 
Moreover, the photospheric component of O~I $\lambda$7773 absorption might be also affected by the residual telluric absorptions. To get a quantitative estimation of this contamination, we introduce a parameter of blending ratio R$_{b}$ which measures the fraction that the line profile of the PHO component of O~I $\lambda$7773 is overlapped by the telluric absorption. This ratio is calculated as  pEW$^{telluric}_{blending}$/pEW$_{PHO}^{O}$, where pEW$^{telluric}_{blending}$ represents the pEW of the region of telluric feature that is blended with the PHO component of O I $\lambda$7773 and pEW$_{PHO}^{O}$ refers to the pEW of the O-PHO. Results of this parameter measured for each spectrum of our sample are reported in the last column of Table 2. Such a blending ratio is overall small for our sample, with a mean value of only 3.7\%, suggesting that the effect of residual telluric lines be negligible. Besides the uncertainties listed above, the PHO component of O~I $\lambda$7773 could also suffer from blending of an unknown component that could be due to Mg~II $\lambda$7890 or additional absorption feature of oxygen at lower velocities (i.e., $<$7,000 km~s$^{-1}$). However, these two effects are not considered in this work because they are difficult to be quantified.

\subsection{Temporal Evolution of O~I $\lambda$7773}
The temporal evolution of the line velocity and strength of O~I~7773 absorption is shown in Figure 3 for some well-observed SNe~Ia, including SNe 2002dj, 2005cf, 2009ig, 2011fe, 2012fr, and 2013dy (see Table 1 for references). According to the spectroscopic classification as proposed by \citet{wa09a}, SN 2002dj and SN 2009ig belong to the HV subclass of SNe~Ia, while SN 2005cf, SN 2011fe, and SN 2013dy can be put into the Normal subclass. SN 2012cg shares some properties similarly seen in the SN 1999aa-like subclass that is characterized by shallow silicon in the spectra \citep{si12b, zh14, marion15}. SN 2012fr may lie at the boundary of the above classifications \citep{zh14}.

As shown in the left panels of Figure 3, the O~I $\lambda$7773 absorption of different SNe~Ia shows large differences in the velocity and the velocity evolution for both the HVF and PHO components. For example, at t$\approx$-10 days, the PHO velocity measured for our sample has a range from $\sim$11,000 km~s$^{-1}$ to $\sim$15,000 km~s$^{-1}$, and the velocity of the HVF-I ranges from $\sim$17,000 km~s$^{-1}$ to $\sim$20,000 km~s$^{-1}$. The O~I line (both PHO and HVF-I) of the HV SNe Ia has a velocity that is on average larger than that of the Normal ones by 3000-4000 km s$^{-1}$ at comparable phases, as is similar to the case seen in Si~II $\lambda$6355 \citep{zhao15}. Restricting the data to the phases from t $\sim$ $-$13 days to t $\sim$ $-$7 days in the calculations of velocity gradient, we find that the velocity gradient measured for the PHO components of SN 2005cf, SN 2011fe, and SN 2013dy is $-$182, $-$186, and $-$100 km~s$^{-1}$~d$^{-1}$, respectively, while the corresponding values obtained for their HVFs are $-$183, $-$227, and $-$263 km~s$^{-1}$~d$^{-1}$. For SN 2002dj, SN 2009ig, SN 2012cg, and SN 2012fr, the velocity gradient of the PHO component is $-$379, $-$199, $-$329, and $-$257 km~s$^{-1}$~d$^{-1}$, while the corresponding value for the HVF is $-$134, $-$644, $-$136, and $-$413 km~s$^{-1}$~d$^{-1}$. The Normal SNe Ia appear to have relatively uniform velocity evolution compared to other subclasses of SNe Ia. This may enable better extrapolations of absorption velocity to the value at a given phase (i.e., t $\sim$ $-$10 days) for some normal sample when necessary. By contrast, the velocity evolution shown by SN 2002dj and SN 2009ig show large scatter either for the HVF or the PHO component, which may be more or less due to that the HV SNe Ia suffer more serious line blending and it is difficult to separate the HVFs from the PHO components.

The strength of the O~I $\lambda$7773 absorption is plotted as a function of phase in the right panels of Figure 3, which shows even larger scatter than the velocity. One can see that the O~I absorptions are found to be strong in some SNe Ia such as SN 2002dj and SN 2011fe but they are marginally detected in other objects such as SN 2009ig and SN 2012fr. Note, however, that the absorption strength of the O-HVF does not show significant variations with time for the our sample, especially the Normal subsample. This is different from the trend seen in the HVFs of Si~II $\lambda$6355 line and Ca~II~NIR triplet, which become weak at a faster pace perhaps due to having optically thin environment in outermost layers \citep[see the references in][]{zhao15}.

\section{Statistical Analysis\label{S_correlation}}
Table 1 lists the line velocities and strengths of O~I $\lambda$7773, Si~II $\lambda$6355, Ca~II~NIR triplet, and C~II $\lambda$6580 absorptions obtained at t $\sim$ $-$10 days for 62 SNe~Ia. If, however, more than one spectra are available during the period from t = $-$11 days to t = $-$9 days, the median values are presented. The B-band light-curve decline rate $\Delta$m$_{15}$(B)(Phillips 1993) is also listed. The detailed results about the measurements of velocities and pEW of the O~I $\lambda$7773 and C~II $\lambda$6580 absorptions at different phases are tabulated in Tables 2 and 3. Given the uncertainties due to line blending and/or intrinsically larger scatter in temporal evolution of O~I absorption for HV SNe Ia, we concentrate on the spectroscopically normal SNe Ia with v$^{Si}_{0}$ $<$ 12,500 km s$^{-1}$ in the following analysis.

\subsection{Expansion Velocity from O~I~7773}
Velocity distribution of the ejected matter at different layers of the exploding WD can provide strong constraints on its compositional structure. To get an overall picture of velocity distribution of C~II $\lambda$6580, Si~II $\lambda$6355, O~I $\lambda$7773, Ca~II~NIR triplet, we construct the mean spectrum using spectra of a subsample of Normal SNe Ia whose spectra have a sufficiently high signal-to-noise (S/N) ratio (i.e, S/N $\gtrsim$ 20) and cover phases close to t = $-$ 10 days. The mean profiles of the above four  absorption features are shown in Figure 4, where the Ca~II lines are characterized by prominent HVFs at $\sim$22,000 km s$^{-1}$ and the O~I line is characterized by the HVF-I at $\sim$18,000 km s$^{-1}$ and the possible HVF-II at $\sim$22,000 km s$^{-1}$. The small notch on the blue side of Si~II $\lambda$6355 absorption corresponds to the HVF of Si formed at a velocity of about 18,000 km s$^{-1}$. The C~II $\lambda$6580 absorption does not show any significant signature of absorption feature at high velocity. As it can be seen, the photospheric components of different species have similar velocities (see blue dashed line in Figure 4). Note that Figure 4 is constructed for illustrative purpose only.

Figure 5 shows the expansion velocities measured from absorption minima of Si~II, Ca~II, and C~II lines versus that from the O~I line.
At the photospheric layer, the velocity of O~I line shows a positive correlation with that of Si~II, Ca~II, and C~II lines; and a linear fit indicates that the O~I velocity is slightly lower than the Si~II and C~II velocities by about 750 km s$^{-1}$. At outer layers, the HVF-I of O (referred as O-HVF for short) shows a similar but slightly lower central velocity compared to the Si-HVF; and at very outer layers, the HVF-II of O has a velocity that is roughly comparable to the Ca-HVFs. This similarity in expansion velocity indicates that the fuel-indicative O and the burned Si (or Ca) are physically connected. The observation results for the small velocity differences between them and that the Ca-HVFs have higher velocities relative to the Si-HVF might be well explained by different ionization (or excitation) energies required to produce lines of O~I $\lambda$7773 (i.e., E$_{ex}$ = 9.15 ev), Si~II $\lambda$6355 (i.e., E$_{ex}$ = 8.12 ev), and Ca~II~NIR triplet (i.e., E$_{ex}$ = 1.7 ev). For a relatively low temperature expected for the outermost HVF-forming layer, Ca~II~NIR triplet are more easily formed and saturated compared to the Si~II and O~I lines.

\subsection{Equivalent Width of O~I~7773}
The line strength of O~I~7773 absorption carries important information on the diversity of SNe Ia. Figures 6 shows the pEW (PHO and HVF) of O~I absorption versus the corresponding velocities, measured from the t $\approx$ $-$10 day spectra of our SN Ia sample. As can be readily seen in these figures, these two quantities do not show significant correlation for the full sample, as indicated by the lower Pearson coefficients. Excluding the HV SNe Ia from the statistical sample, a modest anti-correlation seems to emerge between the pEW and velocity of O~I lines, though there are a few outliers. These results indicate that a smaller amount of oxygen is detected in SNe Ia with relatively larger expansion velocities. Such an inverse relation becomes stronger for the subsample with prominent detection of C~II $\lambda$6580 absorption, i.e., pEW$\gtrsim$ 1.0 \AA\ at t $\approx$ $-$10 days (see table 3 for the detailed results of the measurements). And the corresponding Pearson coefficients $\rho$ are $-$0.64 and $-$0.67 for the correlations of the HVF and PHO components, respectively. The much tighter pEW-velocity relation for O~I $\lambda$7773 absorption suggests that the SNe Ia showing prominent carbon in spectra may form a distinct population with relatively smaller diversity.

We notice that the velocity-pEW correlation seen in O~I $\lambda$7773 absorption is contrary to that seen in Si~II $\lambda$6355 line \citep{wa09a, bl12}, which directly reflects the different roles that O and Si play in the burning in the HVF and photospheric layers. In the rightmost panel of Figure 6, we further examined the relation between the absorption strengths of the HVF and PHO components of O~I $\lambda$7773. We find that the pEW of the O-HVF absorption is highly correlated with that of the O-PHO absorption, which has a Pearson coefficient of 0.86. This strong correlation indicates that the O detected at very outer layers of the ejecta is intrinsic to the SNe, rather than from other sources such as CSM, which usually lies far outside the exploding WD and should not have such a strong connection with the photosphere unless the CSM properties is somehow tied with SN properties. Note that the O-HVF we discussed here refers to the HVF-I marked in Figure 1, and we did not attempt to quantify the correlations of the HVF-II of O with the HVFs of Si (or Ca) because this feature is usually very weak and there are relatively larger uncertainties in measuring its absorption strength.

\subsection{Correlations of O~I $\lambda$7773, Si~II $\lambda$6355, and Ca~II~NIR triplet}\label{S_sica}
In our previous study, HVFs of both Si~II $\lambda$6355 and Ca~II~NIR triplet in SNe~Ia were systematically examined using their early-phase spectra. While the PHO velocity was found to be similar between Si~II $\lambda$6355 and Ca~II~NIR triplet, the HVF velocity of the latter is higher than the former by about 4,000 km~s$^{-1}$. Similarly, although Ca~II~NIR triplet has a PHO strength that is roughly comparable to the Si~II $\lambda$6355, its HVF is found to be much stronger (by about 6 times). Note that these correlations are all positive, meaning that the velocity and strength of these two lines grow/decline in the same direction. These two spectral lines were also found to show similar behaviors as functions of photometric and host-galaxy properties like $\Delta$m$_{15}$($B$), $B_{max} - V_{max}$ color, host-galaxy type etc. \citep[see details in][]{zhao15}. These results indicate that the amount of Si and Ca in both PHO- and HVF-forming regions are strongly connected. Nevertheless, the ratio of Si to Ca is not sensitive to different burning processes expected in SNe~Ia (indeed which is similar to the unburned solar composition), while the ratio of O to Si (or Ca) is expected to be quite different in different layers.

Figure 7 compares the line strengths of O and Si for our sample of SNe Ia. From the left panel of the plot, one can see that the PHO components of O and Si are positively correlated and the Pearson coefficient is 0.69 for the Normal SNe Ia. On average, the pEW of O-PHO absorption is found to be roughly 1/3 times that of the Si~II. Inspection of the right panel of Figure 7, however, reveals that there is a distinct anti-correlation between the absorption strengths of the HVFs of Normal SNe Ia, with the Pearson coefficient $\rho$ = $-$0.71 for a linear relation. Assuming a reciprocal correlation for pEW$^{O}_{HVF}$ - pEW$^{Si}_{HVF}$ relation, the Pearson coefficient becomes 0.90. Similar correlation (for the PHO component) and anti-correlation (for the HVF) can be also found between the line strengths of O~I $\lambda$7773 and Ca~II~NIR triplet lines as shown in Figure 8, though both these relations are less significant. Compared to Normal SNe Ia, the HV ones (i.e., $v^{Si}_{0}$ $\gtrsim$ 12500 km s$^{-1}$) show large scatter in both Figure 7 and Figure 8, and do not seem to follow well the above relations in particular the anti-correlation of the HVFs. The spectroscopically peculiar and luminous objects, with weak absorptions of both O and Si, also show obvious deviations in the plots of the HVFs.

The anti pEW$^{O}_{HVF}$ - pEW$^{Si}_{HVF}$ correlation indicates that, at very outer layers of the exploding WD, less amount of O will be detected when there are more abundant of Si and Ca, and this favors the need for the oxygen burning to produce those HVFs. Again the HV subgroup of SNe Ia are found to show larger scatter in this correlation, suggesting that their HVFs may have different origins (see discussions in \S 4).

\subsection{Correlation of the Absorption with Peak Luminosity}\label{S_dm15}

Since the line velocity and strength of O~I~7773 absorption feature shows a wide range for different SNe Ia, it is necessary to explore the reasons for this diversity. Peak luminosity is an important parameter reflecting the properties of SNe Ia. In Figure 9, the observed features of O-PHO and O-HVF (including both line velocity and absorption strength) are plotted against the luminosity-indicator parameter $\Delta$m$_{15}$($B$). As it is readily seen, the velocity of the O-HVF shows a modest dependence on $\Delta$m$_{15}$($B$), with slower-declining (or more luminous) SNe~Ia having larger ejecta velocities at outer layers. This velocity-luminosity relation can be explained if the characteristic velocity is moved toward higher velocities for explosions with more complete burning. Due to the loss of absorbing oxygen material, the strength of the absorption feature could then be weakened for SNe Ia with higher luminosities. This is supported by the prominent pEW - $\Delta$m$_{15}$(B) relation as shown in the right panels of Figure 9, where stronger O~I absorptions are found in SNe Ia with lower luminosities. On average, the SNe~Ia with $\Delta$m$_{15}$($B$) $\gtrsim$ 1.20 mag have pEWs that are about 2.0 times larger than those with $\Delta$m$_{15}$($B$) $<$ 1.20 mag. The above relations become {\bf stronger} when the HV SNe~Ia are discarded in the analysis.

As a comparison, the correlations of Si-HVF and Ca-HVF (line strength and velocity) with $\Delta$m$_{15}$(B) are also examined in Figure 10. The velocities of Si- and Ca-HVF are also found to be lower for SNe Ia with larger decline rates, which is similar to the behavior shown by the O-HVF. Although the strength of Si-HVF does not show a strong anti-correlation with $\Delta$m$_{15}$(B), it is clear that stronger Si-HVF tends to be detected in more luminous SNe Ia (i.e., with $\Delta$m$_{15}$(B)$<$1.3 mag). This tendency is consistent with the earlier result obtained using the relative line strength of the Si- and Ca-HVFs \citep{ch14, sil15, zhao15}. The correlation between Si-HVF and $\Delta$m$_{15}$(B) is weaker than that observed between O-HVF and $\Delta$m$_{15}$(B), which can be due to that not all luminous SNe Ia have prominent Si-HVF and the line blending of the HVF and the photospheric component is more serious than that for the O~I $\lambda$7773 absorption. The fact that the Si-HVF and O-HVF show an opposite correlation with $\Delta$m$_{15}$ suggests that ionization effect should not play a key role in forming the HVFs in the outer layers of the ejecta since Si and O have similar ionization energy (see detailed discussions in \S 4.1).

\section{Discussions}

\subsection{Origin of the High Velocity Features}\label{S_origin}

The origin of HVFs in SNe~Ia still remains unclear. It has been suggested that HVFs could be associated with abundance enhancement (AE), density enhancement (DE) or ionization enhancement (IE) in the outermost layers \citep{ge04,ma05a,ma05b,bl12}. The material producing the HVFs could be either intrinsic to the SNe or from the CSM. The features we investigate in this work, i.e., O~I $\lambda$7773, Si~II $\lambda$6355 and Ca~II~NIR triplet are suited to probe density structure (through pEW), velocity distribution (see Figure 4), and composition (through pEW) of the ejecta. With these information, we may decode the main functions involved in the origin of some HVFs.

Here we briefly summarize these scenarios so far proposed for the formation of HVFs. In the AE scenario, the abundances of Si and Ca are somehow enhanced in the outermost regions of the ejecta. A possible cause is a strong asymmetry in the explosion process (see Maeda et al. 2010 and Seitenzahl et al. 2013 for SD scenario, or R\"{o}pke et al. 2012 for DD scenario). Alternatively, the abundances could be enhanced by the He-burning near the WD surface, as suggested by double-detonation model \citep[e.g.,][]{fin07,woo11}. In the DE scenario, the HVFs are suggested to generate in a dense shell of basically unburned material, formed either at the outermost layer of the ejecta or CSM \citep{ge04,ma05b,mul15,ta06,ta08}. In the IE scenario, a small amount of H in the outermost layer serve as a source of free electrons, which thus suppresses the ionization status of Ca and Si through recombination. It then leads to a larger amount of Ca~II and Si~II, potentially producing the HVFs \citep{ma05a, ta08}. This may happen either as a contamination of H in the WD surface before the explosion or due to an interaction between the ejecta and the H-rich CSM as is similar to the DE scenario.

In our previous study \citep{zhao15}, the HVF of Si~II $\lambda$6355 is compared with that of Ca~II~NIR which is much stronger (the difference in strength could be understood as coming from different ionization potentials and oscillator strengths of the two lines). Also we found anti-correlation with $\Delta$m$_{15}$($B$), namely slower-declining (or brighter) SNe~Ia tend to show more prominent HVFs. In this work, we further examine the behavior and correlations of the fuel-indicative O~I $\lambda$7773 absorption feature. Here we summarize our findings in relation to the expectations from the DE, AE, and IE scenarios.

(1) Mutual correlations of the velocities of HVFs, and their time evolutions: (a) The velocities of Si-HVF are about 4,000 km~s$^{-1}$ lower than the velocities of the Ca-HVF, i.e., v$^{Si}_{HVF}$ $-$ v$^{Ca}_{HVF}$$\approx$4,000 km~s$^{-1}$ (see Figure 17 in Zhao et al. 2015). This may not support DE scenario of the origin of HVFs. If HVFs (of O, Si and Ca) are generated in a dense-shell (either the outermost layer of the ejecta or CSM), their (central) velocities should be roughly the same, regardless of the radiation condition, unless the shell is very thick in radial scale. But such a large-scale density enhancement is not expected in hydrodynamics. Also, the velocity of Si-HVF is much lower than the expectation from the CSM scenario. As \citet{ta06} pointed out, dense blobs covering the entire photosphere would result in Si~II $\lambda$6355 absorption velocities in excess of 20,000 km~s$^{-1}$ which is however observed only in a few (HV) SNe \citep[see also the discussion in ][]{bl12}. (b) Similarly, the time evolution of the line velocity of O-HVF as seen in Figure 3 may not support the DE scenario as the origin of HVF either. If the HVFs originates from a dense shell created by the SN-CSM
interaction, one would expect nearly a constant velocity of the HVFs as a function of time since the hydrodynamical interaction is
expected to create a geometrically very thin shell in which the velocity variation is at most a few percent (Chevalier 1982). While
this variation in velocity is much larger than that seen in Figure 3 where the typical velocity variation is close to 20\% within one week from the earliest detection.

(2) Additional HVF of O~I $\lambda$7773: as one can see from Figures 1,2, and 4, the HVF of O~I $\lambda$7773 is accompanied by additional HVF at even higher velocity (i.e., higher by $\approx 4,000$ km~s$^{-1}$). Possible explanation for this doublet-HVF includes the followings: {\it Explanation A}-- The O-HVF may be a combination of burned and unburned clumps, where the burned clumps also form the Si- and Ca-HVFs. While the detection of Si- and Ca-HVFs suggests that the burnt clumps are distributed in a large velocity space, at high velocity, the fraction of unburned clumps may be large and then the HVF-II of O~I may be dominated by such unburned ones. Given the lower ionization energy for Ca II, the HVFs can be still formed for Ca~II~NIR triplet at higher velocities where a smaller amount of the
burnt clumps exist, but not for Si II. Alternatively, we have {\it Explanation B}-- The HVF-II of O~I could be produced from carbon burning initiated by outflowing flames at higher velocities \citep{ma05a,mae08} in an asymmetric explosion or from He burning near the WD surface. The burning may also light up HVFs of Ca~II~NIR which has very low excitation energy at early times. However, these processes will not be the dominant factor affecting the correlation with $\Delta$m$_{15}$(B) because the amount of helium or carbon near the WD surface is small. Finally, there is {\it Explanation C} -- The HVF-II of O~I could be from a shocked CSM, while the HVF-I is from the outer layers of the ejecta.

(3) Slow time evolution in the line strength of O-HVF: for the Normal SNe~Ia, the absorption strength of O-HVF weakens at a much slower rate (see Figure 3) than the Si-HVF in early phases (see Figure 5 in Zhao et al. 2015). This result can not be explained by ionization effect because Si~II $\lambda$6355 has an excitation energy even lower than the O~I $\lambda$7773 (i.e., 8.12 ev vs 9.15 ev). Nor would this be easy to explain with the DE scenario. If the fast weakening of Si- and Ca-HVFs is caused by the fast declining of density in the HVF layers, then the O-HVF would also quickly weaken for the same reason. The most plausible explanation could be given by the scenario related to the AE. If the abundance of O increases towards lower velocities, then the decreasing density as a function of time could be compensated by the increasing abundance, resulting possibly in a slow evolution. In other word, if a main body of the HVF-I forming region is dominated by the burnt material and these clumps indeed become less significant for the lower velocities toward the photosphere, then the observed behavior is reproduced.

(4) Velocity-pEW correlation of O~I absorption: as it can be seen from Figure 6, the line strength of O absorption (both HVF and PHO) is decreasing with increasing velocity. This anti-correlation might be explained by the fact that more complete burning of oxygen could release more energy, deriving the remained oxygen shell to move at a higher velocity. The characteristic velocity is then moved toward the higher velocities for explosions with more complete burning. Due to the loss of absorbing oxygen material, the strength of the absorption feature could then be weakened for SNe having higher velocities. This inverse correlation is fully in line of the effect from burning difference. However, it is not clear why this relation is only strong in the SNe~Ia showing a signature of C~II $\lambda$6580 absorption. Possibly, it would indicate that SNe Ia showing the strong C~II would form a distinct population.

(5) Correlation and Anti-correlation between absorption strengths of O-HVF and Si- or Ca-HVFs: from Figures 7 and 8, one can see that there are positive correlations between the photospheric components of O and Si (or Ca) and anti-correlations between their HVFs. These results likely provide an evidence that the HVF and PHO component are created at different characteristic burning layers, which is required in the AE scenario. The HVFs of Si and Ca could be produced from He burning or asymmetric burnings in the outermost layers.
However, this burning process is not expected to have a significant effect on SN Ia luminosity (or $\Delta$m$_{15}$(B)) due to the small amount of He fuel near the WD surface. Observationally, a SN may show significant deviation from the luminosity-$\Delta$m$_{15}$(B) relation established for SNe Ia \citep{phi93} if the helium burning plays an important role. On the other hand, this relation seems difficult to explain in the DE scenario, as the density of O is also enhanced in the density enhancement process. This finding does not directly support the IE scenario (as such an effect is not required in the IE), but nor reject the IE scenario.

(6) Correlation of O-HVF with $\Delta$m$_{15}$($B$): from Figure 9, one can see that the strength of O-HVF tends to become stronger for SNe Ia with larger $\Delta$m$_{15}$($B$). This tendency is opposite to that seen in the Si-HVF (see Figure 10), where stronger HVFs are only detected in SNe~Ia with $\Delta$m$_{15}$($B$)$<$1.3 mag. Indeed, the behavior of the O-HVF is understandable through the general ionization effects. For the SNe~Ia with higher luminosity, their outer materials should be at higher ionization stages. This reduces the number of neutral ions, thus depressing the O~I line -- as is observed. On the other hand, the behavior seen in Si~II is difficult to understand solely from this effect -- they should also be weakened for higher SN luminosity, but the observations indicate the opposite trend. A larger Si/O pEW ratio is observed in our sample for more luminous SNe Ia, a result which solidly rejects the ionization effect as a possible dominant factor.

(7) Abnormal behavior of the HV SNe Ia: note, however, the above conclusion may only apply to the SNe~Ia with relatively lower expansion velocities (i.e., v$^{Si}_{0}$ $<$ 12,500 km s$^{-1}$) since the HV SNe~Ia are found to show significant scatter in the mutual correlations and anti-correlations of line strengths between O and Si (or Ca). Large scatter is also seen in the pEW - $\Delta$m$_{15}$($B$) correlations (see Figure 9). Thus, additional mechanism may be needed to explain the formation of the HVFs seen in HV SNe~Ia if the measurements of their HVFs are generic. Given that the HVFs of Si in HV SNe have higher velocities than those in Normal SNe, a possible explanation is that the burning effect is weakened as the shells move outward. As an alternative, the HVFs of HV SNe Ia might also arise from the density enhancement of outer Si shell, perhaps due to the CSM interaction. Also, we note that indeed the HV SNe Ia could come from multiple populations, one belonging intrinsically to the same population as Normal SNe \citep{mae10} and the other exploding in the younger environment than Normal SNe \citep{wa13}. In this case, it could be natural that the HV SNe show diversity in the HVFs as well. Nevertheless, there is one caveat that the scatter in the HV SNe Ia could merely come from the uncertainty in the fitting.

In conclusion, the different behaviors of HVFs of O and Si (or Ca), especially the anti-correlations between the HVF strengths, are most naturally explained by a scenario where the HVF regions experienced explosive oxygen burning. This result is consistent with, and indeed expected for, the AE scenario.

\subsection{Constraints on Explosion Models}
Besides clarifying the origin of the HVFs formed at the outermost layers of the exploding WD, our result also places a strong constraint on the still-debated explosion mechanism. The need for formation of the HVFs from nuclear burning leaves us three possible models: standard delayed detonation \citep{kh91,gam05}, double detonation through He accretion \citep{fin10,woo11}, and violet merger of two WDs \citep{pak11, ro12, pak12}.

It has been suggested that the double detonation model will produce mostly the Fe-peak elements at high velocity \citep{woo11} rather than intermediate mass elements (IMEs), and the C and O will not be there in the He layer. Therefore, this model is not good to explain the HVFs seen at very outer layers of the ejecta. In the case of violet mergers, the detonation can convert the bulk of the secondary WD to the IMEs. The simulation indicates that these IMEs will indeed be left in the low-velocity zone of the ejecta, with typical velocities $<$ 20,000 km s$^{-1}$ \citep{ro12}, and no mechanism proposed so far can accelerate the IME-rich region toward the higher velocity in this scenario. In addition, for models of either double denotation or violet merger, the "surface" detonation and the resultant HVFs are controlled by the nature of the mass accretion and/or secondary star, while the main features of the SN Ia will be determined by the primary WD. Therefore, it is difficult to understand the luminosity-velocity relation shown in Figures 9 and 10 within the framework of these two models.

On the other hand, the delayed-detonation model has a natural explanation for the observed brighter-faster relation. The diversity of the outer-layer spectral features can be attributed to the difference in the transition density $\rho_{tr}$ from deflagration to detonation in the explosion \citep{ho02, woo09}. In case this transition is delayed, the expansion velocities of the ejecta will decrease and the burned materials like Si (or Ca) will be less abundant at higher velocities because a significant amount of oxygen remains unburned and does not contribute to the energy production. This supports the notion that the degree of burning is an important source of spectroscopic diversity among SNe Ia in addition to the progenitor scenarios.

In the currently available MULTI-DIMENSIONAL simulations for delayed detonation models, the velocity is still limited to $\sim$20,000 km s$^{-1}$, but this velocity may be further extended once the outermost region is well resolved, and thus the higher resolution simulations may show small clumps (currently not resolved) penetrating into the outermost layer. The observed velocity distribution of O and Si and the Si/O ratio in the outer layers of the ejecta may be used to constrain the density where the transition from deflagration to detonation occurs by comparing with the predictions from models \citep{ho02, sei13}. However, high-resolution simulations are further needed to provide a guide to connect the detailed hydrodynamic nature of the explosion and the observed properties we have found in this paper.

\section{Conclusion}\label{S_conclusion}

With a large sample of early spectra (t$\leq$ $-$7 days), we search for the HVFs in the absorption features O~I $\lambda$7773, Si~II $\lambda$6355 and Ca~II~NIR triplet. Double O-HVFs are detected in early-time spectra of SNe Ia, with velocities comparable to those of the Si-HVF and Ca-HVFs, respectively. Their mutual correlations and correlations with $\Delta$m$_{15}$($B$) are scrutinized.

By comparing the HVFs of O~I with those of Si~II and Ca~II, we attempt to differentiate between various scenarios on the formations of HVFs at outer layers of the exploding ejecta. From the anti-correlation between the pEWs of HVF of O~I and those of Si~II and Ca~II, we conclude that the oxygen burning is an important contributor to the HVFs of Si and Ca at least for Normal SNe~Ia (see discussion in \S \ref{S_origin}). This evidence is against the scenario that Si/Ca HVFs are produced from primordial material (i.e., CSM), while it is in line with the abundance enhancement scenario. Considering that the HV SNe~Ia tend to have distinct explosion environments \citep{wa13, zhao15} and weaker correlations and anti-correlations between O~I and Si~II (or Ca~II), it is possible that the formation of their HVFs is more complicated than the Normal counterparts -- indeed the HV subclass may come from multiple populations either having a normal-velocity counterpart or not \citep{mae10, wa13}, so this could complicate the analysis of the HV SNe Ia. Given that the HVFs of Si in HV SNe Ia have higher velocities than those in Normal SNe Ia, a possible explanation is that the burning effect is weakened as the shells move outward. As an alternative, the HVFs of HV SNe Ia might also arise from the density enhancement of outer Si shell, perhaps due to the CSM interaction.

Besides distinguishing the origin of HVFs, the velocities and strengths we have measured for species of C, O, Si and Ca in this paper could be used to shape a picture of the ejecta, and further constrain the explosion models. The existence of HVF-I and even HVF-II of O~I $\lambda$7773 at higher velocities indicates, the photosphere of SNe~Ia is covered by oxygen materials (clumps or a separated shell) from the white dwarf and explosive carbon burning. Current observations suggest that the delayed-detonation is the favorable explosion model for at least spectroscopically normal SNe Ia with normal photospheric velocities. Numerical explosion simulations with sufficient resolution are encouraged to focus on the outermost layer to further discriminate the explosion models using the new observational indicators we have found in this paper.

\acknowledgments We thank the anonymous referee for his/her suggestive comments to help improve the manuscript. We are grateful to the 
staffs of the various telescopes and observatories with which data were obtained. The work is finally supported by the Major State Basic Research Development Program (2013CB834903), the National Natural Science Foundation of China (NSFC grants 11178003, 11325313, 11403096, and 11203034), the Strategic Priority Research Program ``The Emergence of Cosmological Structures" of the Chinese Academy of Sciences (grant No. XDB09000000), and the China Scholarship Council (CSC 201406210312). The work by K.M. is partly supported by JSPS Grant-in-Aid for Scientific Research (No. 26800100) and by World Premier International Research Center Initiative (WPI Initiative), MEXT, Japan. This research has made use of the CfA Supernova Archive, which is funded in part by the US National Science Foundation through grant AST 0907903. This research has also made use of the Lick Supernova Archive, which is funded in part by the US National Science Foundation.
This work was also partially Supported by the Open Project Program of the Key Laboratory of Optical Astronomy, National Astronomical 
Observatories, Chinese Academy of Sciences. Funding for the LJ 2.4-m telescope has been provided by CAS and the People's Government 
of Yunnan Province.

\clearpage

\begin{figure*}
\epsscale{.95} \plotone{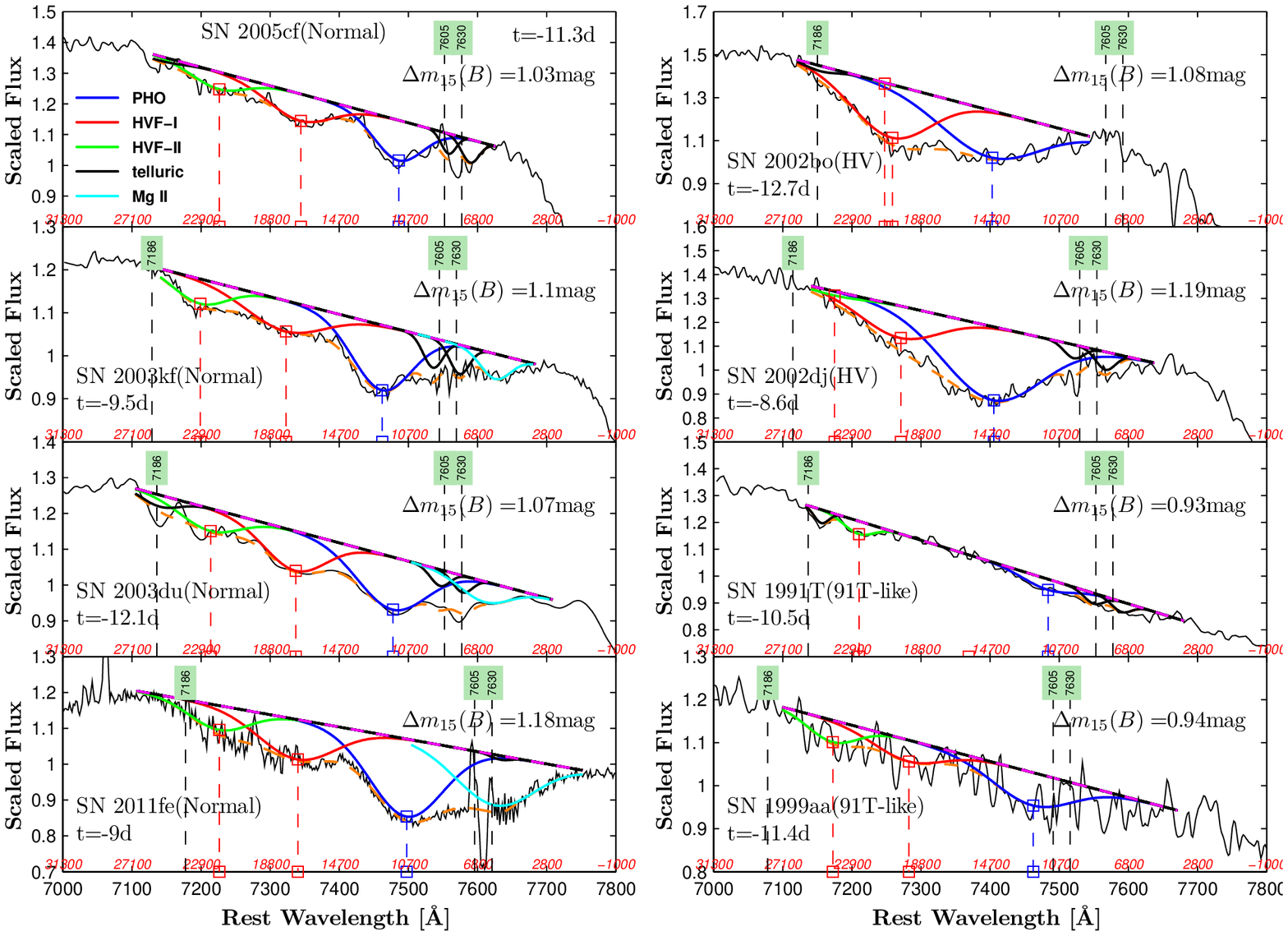}
\caption{\label{Fig1} The absorption features of blueshifted O~I $\lambda$7773, centered near 7300-7500 \AA~ in the optical spectra,
are shown for some representative SNe Ia. The SN name and the phase of the spectrum are shown in each panel. The left panels show the
Normal SNe Ia in order of increasing strength (from top to the bottom) for O~I $\lambda$7773 absorption, while the right panels show the HV and 91T/99aa-like SNe~Ia. Multiple gaussian fit is applied to determine different absorption components, with the blue curve
representing the photospheric component, the red curve representing the main HVF (dubbed as HVF-I), and the green curve representing the the additional HVF at higher velocities (dubbed as HVF-II). The orange-color curve represents the combined fits. The vertical dashed lines mark the positions of absorption minima of the photospheric components and two HVF components. The black dashed lines mark the positions of telluric absorptions (corrected for the redshift of the supernovae) at 7186\AA, 7605\AA, and 7630\AA. The number in the bracket represents the B-band magnitude decline rate over the first 15 days after the maximum light (see text for references).}
\end{figure*}

\begin{figure*}
\epsscale{.95} \plotone{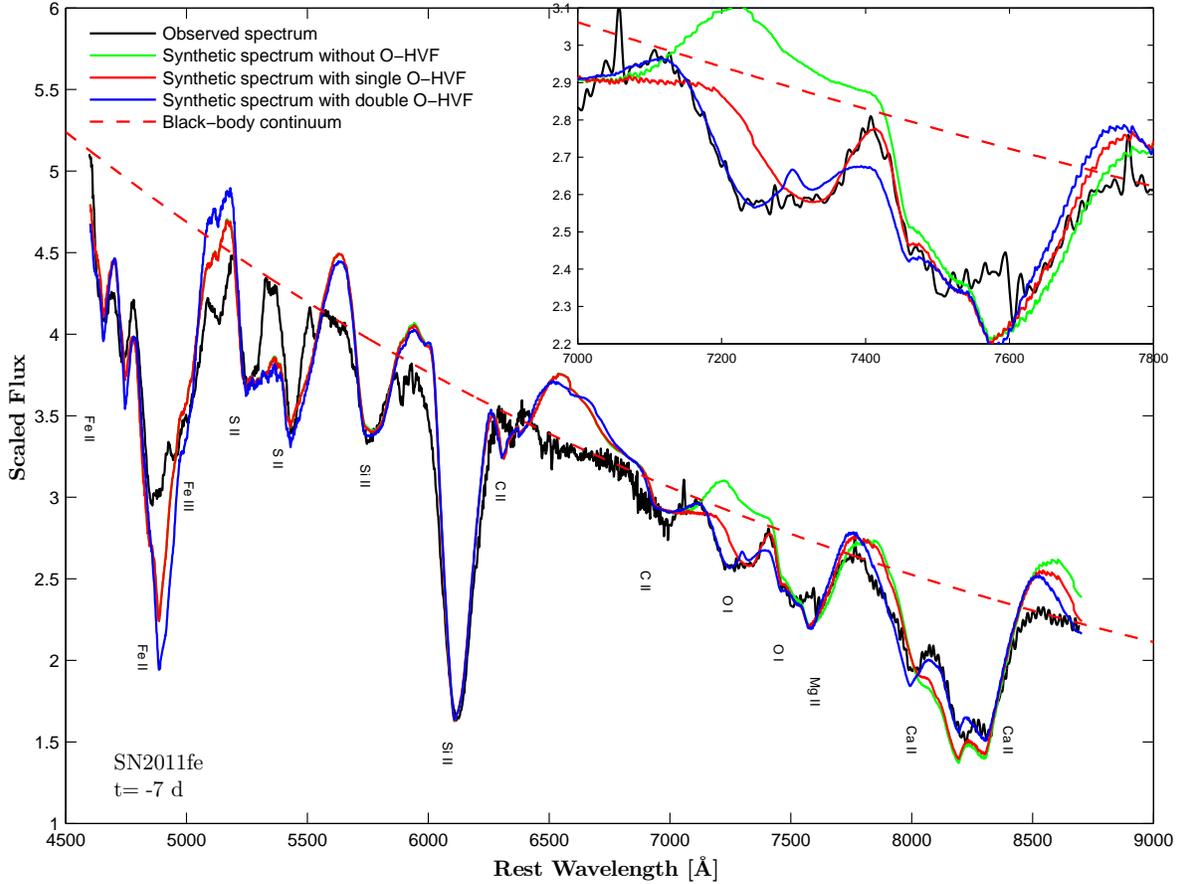}
\caption{\label{Fig2}
The t = $-$7 day spectrum of SN 2011fe is compared with the synthetic spectrum from the SYNOW fit \citep{fish95}. The green curve represents the synthetic spectrum without including the O-HVF; the red curve represents the fit with single O-HVF; the blue curve represents the fit with double O-HVFs at 18,000 km s$^{-1}$ and 22,000 km s$^{-1}$, respectively.}\end{figure*}

\begin{figure*}[!htbp]
\epsscale{.95} \plotone{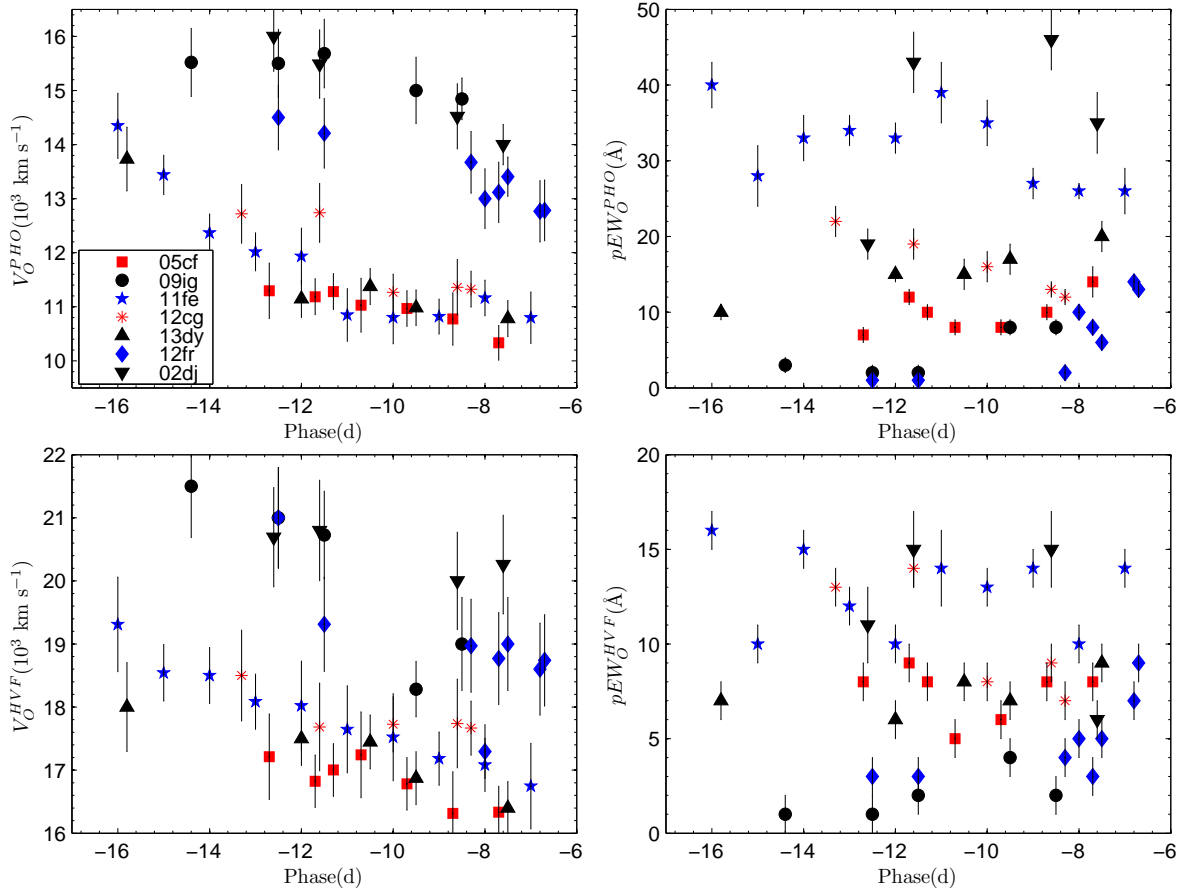}
\caption{\label{Fig3} Left panels: Evolution of the absorption strength of HVF and PHO component of O~I $\lambda$7773 for some well-observed SNe Ia such as SNe 2002dj, 2005cf, 2009ig, 2011fe, 2012cg, 2012fr and 2013dy. Right panels: Similar evolutions but for the absorption strength of the HVF and PHO components of O~I line.}
\end{figure*}

\begin{figure*}[!htbp]
\epsscale{.65} \plotone{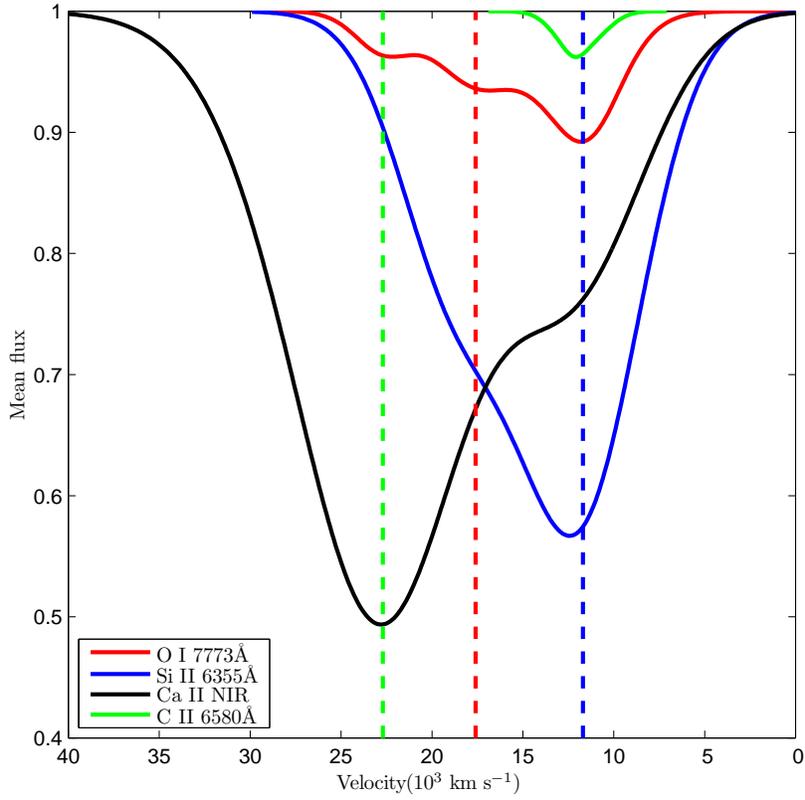}
\caption{\label{Fig4} The mean line profiles of C~II $\lambda$6580 (cyan), Si~II $\lambda$6355 (blue), O~I $\lambda$7773 (red),
and Ca~II~NIR triplet (black) absorption features, obtained with the t$\sim$$-$10 day spectra of spectroscopically normal SNe Ia,
are shown in velocity space. The vertical dashed line marks the position of the absorption minima of HVFs and photospheric component.}
\end{figure*}

\begin{figure*}
\epsscale{.95} \plotone{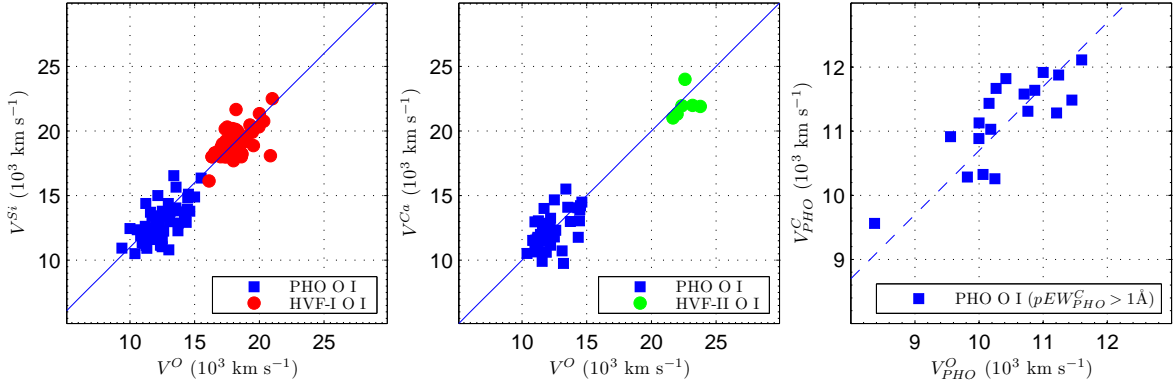}
\caption{\label{Fig5} Comparison of velocity inferred from O~I line with those from Si~II, Ca~II, and C~II lines. The blue dots represent the velocity of the photospheric components. The red dots show the Si-HVF and O-HVF-I, while the green dots show the Ca-HVF and the O-HVF-II. Only those with significant detection of carbon signature are used for the comparison with the C~II velocity.}
\end{figure*}

\begin{figure*}
\epsscale{.95} \plotone{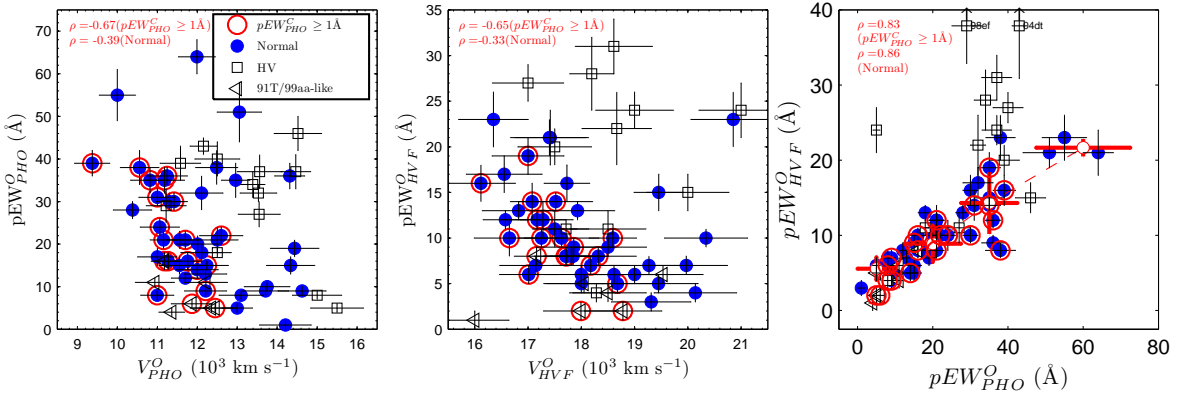}
\caption{\label{Fig6} Line strength and velocity of O~I $\lambda$7773 (a) Comparison between the line strength of the photospheric component of O~I $\lambda$7773 absorption and the corresponding line velocity. (b) Similar comparison as (a) but for the line strength and velocity of the O-HVF. c) Comparison of the line strengths between the HVF and the PHO component. Blue dots show normal SNe Ia with v$^{Si}_{0}$ $<$ 12,500 km s$^{-1}$ at maximum light. The high-velocity SNe Ia ($v^{Si}_{0}$ $\gtrsim$ 12,500 km s$^{-1}$ at maximum light) and the 91T/99aa-like SNe Ia are represented with squares and triangles, respectively. The larger red open circles represent the subsample of SNe Ia showing prominent C~II $\lambda6580$ absorption in the early-time spectra, i.e., pEW $>$ 1.0\AA\ at t$\sim$$-$10 days.}
\end{figure*}

\begin{figure*}[!htbp]
\epsscale{.95} \plotone{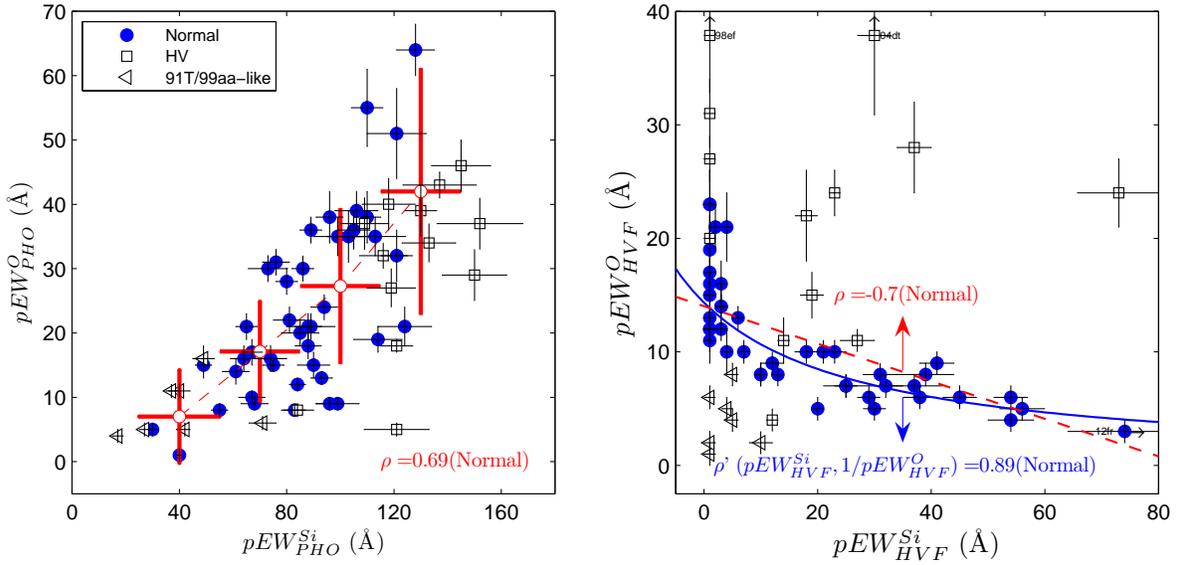}
\caption{\label{Fig7} Comparison of line strength of O~I $\lambda$7773 with that of Si~II $\lambda$6355, with the left panel for the photospheric component and right panel for the HVF. The symbols are the same as in Figure 6. It is clear the PHO components of these two features show a positive correlation (with the Pearson coefficient $\rho$ = 0.69) and their HVFs have an inverse correlation (with $\rho$ = $-$0.70 for a linear correlation and $\rho$ = 0.89 for a reciprocal correlation) for the subgroup of Normal SNe Ia. The open circles in the left panel represent the mean pEWs of O~I $\lambda$7773 in bins of pEWs of Si~II $\lambda$6355, and the error bars are the width of the bins and 1$\sigma$ dispersion.}
\end{figure*}

\begin{figure*}
\epsscale{.95} \plotone{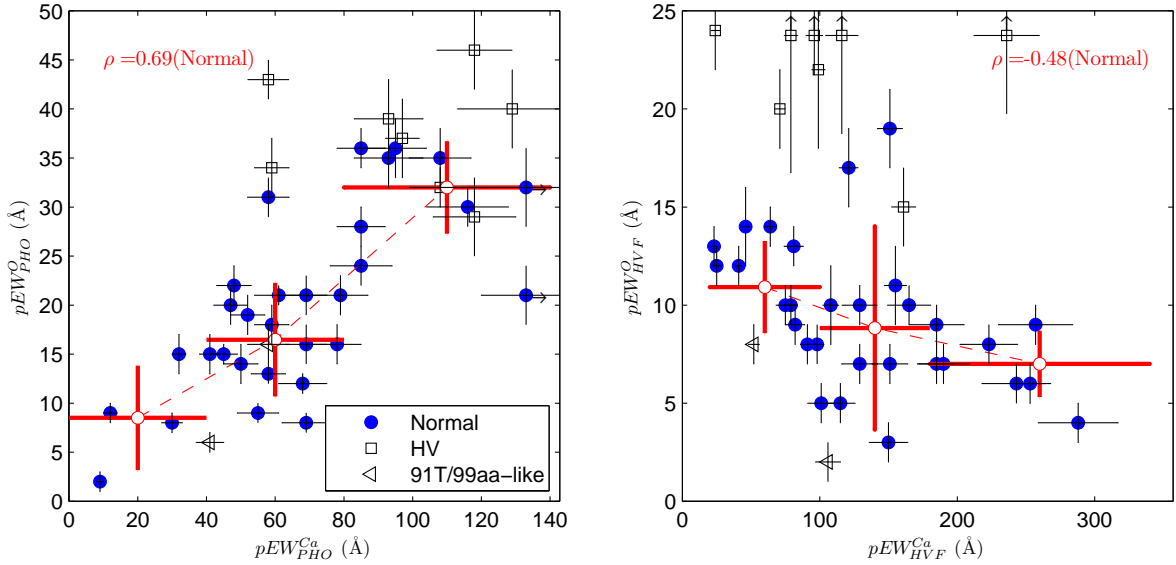} \caption{\label{Fig8} Same as Figure 7 but for the comparison with Ca~II~NIR triplet. Similar  correlation of the PHO component and the anti-correlations of the HVFs exist between O and Ca, but are less significant compared to those seen between O and Si. The Pearson coefficients are 0.69 for the PHO-component correlation and $-$0.48 for the HVF anti-correlation. The symbols are the same as indicated in Figure 6.}
\end{figure*}

\begin{figure*}
\epsscale{.95} \plotone{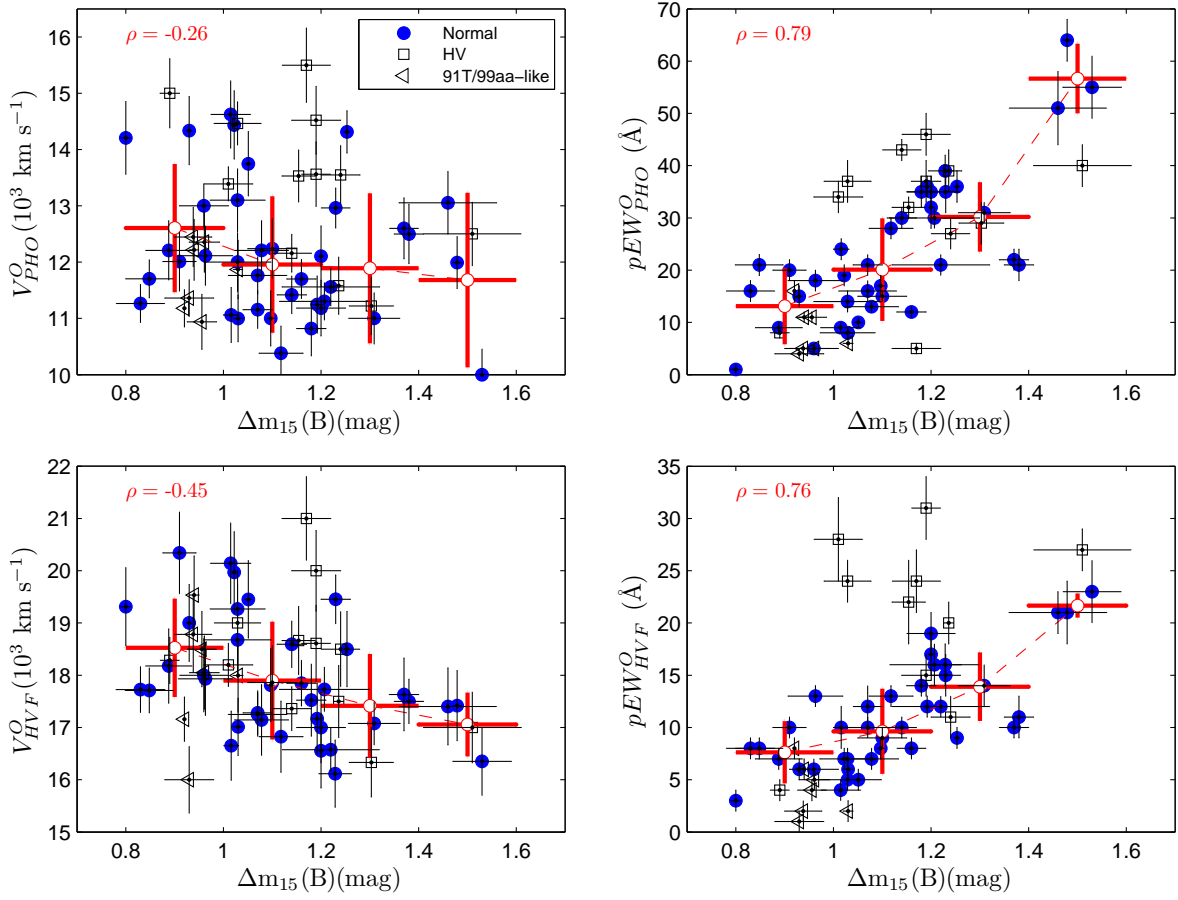} \caption{\label{Fig9} Absorption feature of O~I $\lambda$7773 as a function of the luminosity indicator $\Delta$m$_{15}$(B) shown for both the PHO and the HVF components. It shows that SNe Ia with smaller $\Delta$m$_{15}$(B) (or higher luminosity) tend to have larger expansion velocities (left panels) but weaker absorptions (right panels). For the PHO component (upper panels), the Pearson coefficients of these relations derived for Normal SNe Ia are $-$0.26 for velocity and 0.79 for line strength. For the HVF component, the corresponding coefficients are $-$0.45 and 0.76 for velocity and line strength, respectively. Red circles represent the mean velocity and pEWs of O~I absorption in bins of $\Delta$m$_{15}$(B), and the error bars represent the width of the bins and 1$\sigma$ dispersion. The symbols are the same as indicated in Figure 6.}
\end{figure*}

\begin{figure*}
\epsscale{.95} \plotone{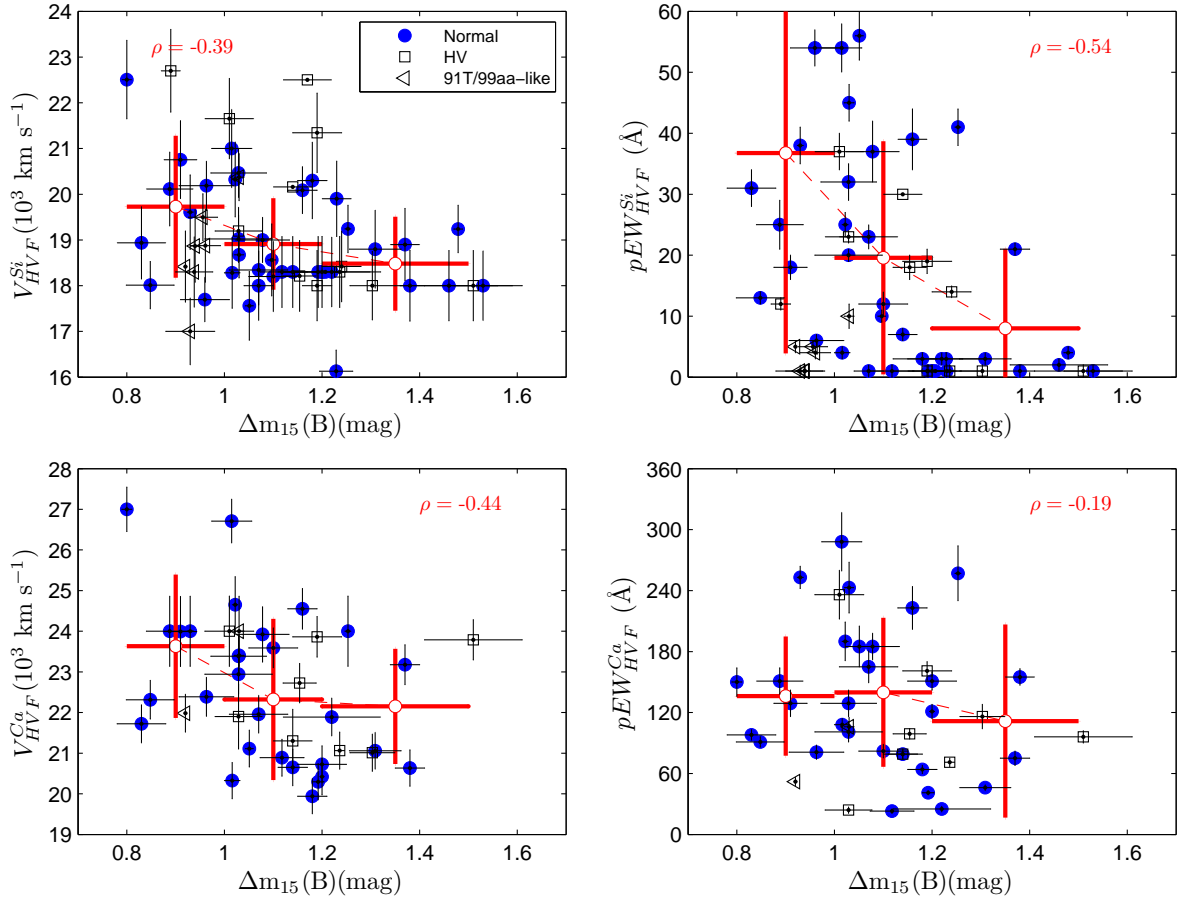} \caption{\label{Fig10} Same as Figure 9 but for correlations of the HVFs of Si~II (upper panels) and Ca~II (lower panels) with $\Delta$m$_{15}$(B).}
\end{figure*}

\clearpage

\begin{deluxetable}{llcccccccccccccccccccccl}
\centering\tabletypesize{\scriptsize}\setlength{\tabcolsep}{0.0002in}\rotate{}
\tablecaption{Velocities and pEWs near t$\sim$$-$10
days\label{T_near-11d}} \tablewidth{0pt} \tablehead{\colhead{}
&\multicolumn{7}{c}{O I $\lambda$ 7773} &\colhead{} &\multicolumn{4}{c}{Si II $\lambda$ 6355} &\colhead{} &\multicolumn{4}{c}{Ca II NIR} \\
\cline{2-8} \cline{10-13} \cline{15-18} \\
\colhead{SN} &\colhead{t\tablenotemark{a}} &\colhead{$V_{P}$\tablenotemark{b}} &\colhead{$W_{P}$\tablenotemark{c}}
&\colhead{$V_{H}$\tablenotemark{d}} &\colhead{$W_{H}$\tablenotemark{e}} &\colhead{$V_{S}$\tablenotemark{f}}
&\colhead{$W_{S}$\tablenotemark{g}} &\colhead{} &\colhead{$V_{P}$\tablenotemark{h}} &\colhead{$W_{P}$\tablenotemark{i}}
&\colhead{$V_{H}$\tablenotemark{j}} &\colhead{$W_{H}$\tablenotemark{k}} &\colhead{} &\colhead{$V_{P}$\tablenotemark{l}}
&\colhead{$W_{P}$\tablenotemark{m}} &\colhead{$V_{H}$\tablenotemark{n}} &\colhead{$W_{H}$\tablenotemark{o}} &\colhead{}
&\colhead{$V_{0}^{Si}$\tablenotemark{p}} &\colhead{$\Delta$m$_{15}(B)$\tablenotemark{q}} &\colhead{Ref.\tablenotemark{r}}

\\
  \colhead{} &\colhead{~~~d~~~} &\colhead{km s$^{-1}$} &\colhead{\AA} &\colhead{km s$^{-1}$} &\colhead{\AA} &\colhead{km s$^{-1}$} &\colhead{\AA} &\colhead{} &\colhead{km s$^{-1}$} &\colhead{\AA} &\colhead{km s$^{-1}$} &\colhead{\AA}
 &\colhead{} &\colhead{km s$^{-1}$} &\colhead{\AA} &\colhead{km s$^{-1}$} &\colhead{\AA} &\colhead{}
 &\colhead{km s$^{-1}$} &\colhead{(mag)} &\colhead{}}

\startdata
1991M & -7.8 & 12501(562) & 40(4) & 17000(675) & 27(2) & 22969(841) & 3(1) & ~ & 13384(624) & 118(10) & 18000(766) & 1(1) & ~ & 14672(345) & 129(16) & 23788(495) & 96(6) & ~ & 12511(434) & 1.51(10) & A;B;11 \\
1991T$^*$ & -10.5 & 11362(331) & 4(1) & 16000(643) & 1(1) & 22539(504) & 2(1) & ~ & 12500(582) & 17(2) & 17000(735) & 1(1) & ~ & \nodata & \nodata & \nodata & \nodata & ~ & 9509(476) & 0.93(05) & A;B;1 \\ 1994D & -9.5 & 12601(542) & 22(2) & 17631(697) & 10(1) & 21679(822) & 5(1) & ~ & 12249(578) & 81(6) & 18900(788) & 21(1) & ~ & 12322(322) & 48(5) & 23177(491) & 75(7) & ~ & 11021(330) & 1.37(03) & A;B;1 \\
1997bq & -11.8 & 13390(302) & 34(3) & 18197(412) & 28(4) & 22603(491) & 14(2) & ~ & 16545(722) & 133(10) & 21654(878) & 37(3) & ~ & 15500(605) & 59(5) & 24000(863) & 236(24) & ~ & 13000(650) & 1.01(05) & A;6 \\ 1998dm & -10.9 & 11702(338) & 21(2) & 17709(429) & 8(1) & 22237(501) & 5(1) & ~ & 11715(356) & 65(4) & 18011(516) & 13(1) & ~ & 12179(319) & 69(6) & 22314(475) & 91(5) & ~ & 11058(553) & 0.85(05) & A;B;8 \\
1998ef & -8.9 & 11222(509) & 29(4) & 16333(664) & 50(5) & 21660(831) & 14(1) & ~ & 14394(634) & 150(12) & 18000(748) & 1(1) & ~ & 11778(314) & 118(12) & 21011(457) & 116(12) & ~ & 13260(384) & 1.30(05) & A;B;12 \\ 1999aa$^*$ & -11.4 & 12216(539) & 11(1) & 19535(750) & 6(1) & 24103(889) & 4(1) & ~ & 12457(590) & 40(4) & 18869(791) & 1(1) & ~ & \nodata & \nodata & \nodata & \nodata & ~ & 10484(310) & 0.94(01) & A;B;2 \\
1999dk & -7.7 & 13531(460) & 32(2) & 18664(651) & 22(4) & 23000(856) & 1(1) & ~ & 13818(637) & 116(9) & 18214(772) & 18(2) & ~ & 14104(352) & 108(9) & 22725(484) & 99(5) & ~ & 12500(625) & 1.15(04) & B;8 \\ 1999dq$^*$ & -9.6 & 12355(531) & 5(1) & 18051(699) & 5(1) & 23586(866) & 5(1) & ~ & 11156(552) & 27(2) & 18875(798) & 4(1) & ~ & \nodata & \nodata & \nodata & \nodata & ~ & 10939(314) & 0.96(04) & A;12 \\
1999ee & -9.4 & 13000(383) & 5(1) & 18000(706) & 6(1) & 23088(509) & 5(1) & ~ & 10800(544) & 30(2) & 17694(490) & 54(3) & ~ & \nodata & \nodata & \nodata & \nodata & ~ & 10189(332) & 0.96(05) & A;B;10 \\ 2000dx$^+$ & -9.0 & 12501(349) & 18(1) & 17715(431) & 11(1) & 22738(507) & 7(1) & ~ & 13793(414) & 121(6) & 19504(520) & 27(3) & ~ & \nodata & \nodata & \nodata & \nodata & ~ & 12500(625) & \nodata & B; \\
2000fa & -9.1 & 12011(524) & 20(2) & 20342(779) & 10(1) & 26000(948) & 1(1) & ~ & 13391(624) & 85(7) & 20756(851) & 18(2) & ~ & 12863(330) & 47(5) & 23994(862) & 129(13) & ~ & 12106(344) & 0.91(04) & A;B;8 \\ 2001el$^+$ & -9.0 & 11701(347) & 12(1) & 17847(437) & 8(1) & 22760(506) & 4(1) & ~ & 13428(368) & 84(3) & 20088(511) & 39(5) & ~ & 14003(338) & 68(7) & 24553(499) & 223(21) & ~ & 11715(346) & 1.16(03) & 10 \\
2002cr & -7.6 & 9371(439) & 39(3) & 16117(644) & 16(2) & 20926(792) & 2(1) & ~ & 10929(328) & 106(8) & 16128(466) & 3(1) & ~ & \nodata & \nodata & \nodata & \nodata & ~ & 10017(292) & 1.23(04) & A;B;11 \\ 2002cs$^+$ & -9.0 & 14465(383) & 37(4) & 19000(736) & 24(2) & 23782(523) & 11(2) & ~ & 14816(446) & 109(9) & 19197(516) & 23(1) & ~ & 13911(346) & 97(5) & 21900(799) & 24(3) & ~ & 13982(700) & 1.03(05) & A;B;8 \\
2002dj & -8.6 & 14521(602) & 46(4) & 20000(771) & 15(2) & 24000(887) & 1(1) & ~ & 15108(677) & 145(11) & 21344(868) & 19(2) & ~ & 14222(351) & 118(11) & 23862(500) & 161(9) & ~ & 13095(655) & 1.19(05) & A;B;1 \\ 2002er & -8.9 & 12962(355) & 35(4) & 19454(465) & 15(2) & 23047(512) & 20(2) & ~ & 14390(654) & 103(8) & 19900(824) & 1(1) & ~ & \nodata & \nodata & \nodata & \nodata & ~ & 11895(346) & 1.23(03) & A;B;1 \\
2003U & -8.8 & 13054(555) & 51(7) & 17403(738) & 21(2) & 24105(891) & 1(1) & ~ & 12985(388) & 121(11) & 18000(766) & 2(1) & ~ & \nodata & \nodata & \nodata & \nodata & ~ & 11046(342) & 1.46(10) & A;B;1 \\ 2003cg & -8.4 & 11559(342) & 21(2) & 16576(420) & 12(1) & 21848(494) & 5(1) & ~ & 11603(350) & 89(9) & 18300(765) & 3(1) & ~ & 12000(497) & 79(8) & 21890(471) & 25(3) & ~ & 11322(326) & 1.22(10) & A;10 \\
2003du & -12.1 & 11762(336) & 16(2) & 17250(424) & 10(1) & 22363(502) & 6(1) & ~ & 12000(581) & 74(6) & 18344(774) & 23(2) & ~ & 11466(304) & 69(6) & 21957(470) & 165(16) & ~ & 10557(306) & 1.07(06) & A;B;1 \\ 2003ek$^+$ & -10.0 & 12483(536) & 38(4) & 20853(794) & 23(3) & 26000(948) & 22(2) & ~ & 11042(330) & 96(5) & 18082(771) & 1(1) & ~ & \nodata & \nodata & \nodata & \nodata & ~ & 10832(320) & \nodata & A;B; \\
2003fa$^*$ & -9.2 & 10938(488) & 11(1) & 18500(719) & 4(1) & 22412(838) & 4(1) & ~ & 11369(336) & 37(2) & 19497(513) & 5(1) & ~ & \nodata & \nodata & \nodata & \nodata & ~ & 10357(518) & 0.96(03) & A;B;4 \\ 2003kc & -11.4 & 13563(573) & 37(4) & 18611(725) & 31(3) & 21409(807) & 3(1) & ~ & 15646(694) & 152(16) & 18000(767) & 1(1) & ~ & \nodata & \nodata & \nodata & \nodata & ~ & 12723(608) & 1.19(03) & A;B;12 \\
2003kf & -9.5 & 12240(535) & 15(1) & 17860(710) & 9(1) & 23000(856) & 5(1) & ~ & 11998(358) & 75(4) & 18200(771) & 12(2) & ~ & 12340(323) & 45(4) & 23590(496) & 82(7) & ~ & 11444(573) & 1.10(05) & A;B;1 \\ 2004dt & -10.8 & 12157(335) & 43(2) & 17362(418) & 64(7) & 21983(494) & 28(3) & ~ & 15000(750) & 137(14) & 20162(1009) & 30(3) & ~ & 12500(512) & 58(6) & 21300(780) & 79(5) & ~ & 14110(406) & 1.14(04) & A;B;C;12 \\
2004eo & -11.4 & 12500(528) & 21(3) & 17500(423) & 11(2) & 24479(502) & 1(1) & ~ & 11684(364) & 124(10) & 18000(774) & 1(1) & ~ & 11800(482) & 144(13) & 20634(445) & 155(8) & ~ & 10497(296) & 1.38(03) & A;B;C;1 \\ 2004ey & -8.2 & 12114(523) & 18(2) & 17930(698) & 13(1) & 23000(856) & 2(1) & ~ & 11759(352) & 88(5) & 20188(532) & 6(1) & ~ & 11474(308) & 59(5) & 22386(477) & 81(7) & ~ & 11220(324) & 0.96(06) & B;C;11 \\
2005cf & -10.2 & 10999(414) & 8(1) & 17012(549) & 6(1) & 22700(678) & 3(1) & ~ & 11773(348) & 83(4) & 18674(503) & 45(3) & ~ & 12963(520) & 69(7) & 23406(492) & 243(25) & ~ & 10276(302) & 1.03(01) & A;B;7 \\ 2005cg$^+$ & -9.0 & 11555(509) & 15(2) & 18000(704) & 5(1) & 22210(831) & 2(1) & ~ & 12137(366) & 49(3) & 18661(806) & 30(2) & ~ & 9906(283) & 32(2) & 23239(489) & 115(11) & ~ & 11590(342) & \nodata & A; \\
2005el & -8.1 & 11000(460) & 31(2) & 17078(416) & 14(2) & 22511(503) & 4(1) & ~ & 11355(338) & 76(5) & 18799(857) & 3(1) & ~ & 11357(304) & 58(6) & 21059(454) & 46(3) & ~ & 10811(304) & 1.31(06) & A;B;C;11 \\ 2005eu$^*$ & -9.1 & 12447(535) & 5(1) & 18782(728) & 2(1) & 23200(862) & 4(1) & ~ & 11161(324) & 42(2) & 18300(774) & 1(1) & ~ & \nodata & \nodata & \nodata & \nodata & ~ & 11068(554) & 0.94(04) & A;B;11 \\
2006X & -11.1 & 15500(660) & 5(1) & 21000(798) & 24(3) & 24040(527) & 7(1) & ~ & 16357(818) & 121(13) & 22500(1125) & 73(8) & ~ & \nodata & \nodata & \nodata & \nodata & ~ & 15577(441) & 1.17(05) & A;B;C;7 \\ 2006ax$^+$ & -10.6 & 11061(490) & 24(2) & 16651(662) & 10(2) & 22137(832) & 2(1) & ~ & 12105(361) & 94(5) & 18281(774) & 4(1) & ~ & 11554(311) & 85(9) & 20324(446) & 108(6) & ~ & 10503(298) & 1.02(02) & A;B;C;3 \\
2006dd & -12.4 & 11156(341) & 21(2) & 17280(427) & 12(2) & 22309(502) & 6(1) & ~ & 11306(340) & 88(5) & 18000(765) & 1(1) & ~ & \nodata & \nodata & \nodata & \nodata & ~ & 10500(525) & 1.07(03) & C;11 \\ 2006dm & -7.9 & 10000(460) & 55(6) & 16351(653) & 23(3) & 21000(795) & 7(1) & ~ & 12454(372) & 110(6) & 18000(765) & 1(1) & ~ & \nodata & \nodata & \nodata & \nodata & ~ & 11800(590) & 1.53(06) & B;7 \\
2006dy & -11.8 & 10556(491) & 38(4) & 18317(722) & 8(1) & 22949(842) & 2(1) & ~ & 12363(370) & 110(5) & 20000(827) & 10(1) & ~ & \nodata & \nodata & \nodata & \nodata & ~ & 10459(523) & \nodata & B; \\ 2006gr & -9.2 & 13100(556) & 8(1) & 19269(744) & 7(1) & 23148(860) & 3(1) & ~ & 13752(418) & 55(3) & 20465(533) & 32(3) & ~ & 10723(299) & 30(3) & 23385(492) & 129(13) & ~ & 11492(366) & 1.03(06) & A;B;4 \\
2006kf & -8.3 & 11991(460) & 64(4) & 17418(673) & 21(3) & 22152(830) & 6(1) & ~ & 12953(390) & 128(7) & 19239(518) & 4(1) & ~ & \nodata & \nodata & \nodata & \nodata & ~ & 11378(322) & 1.48(01) & A;B;C;12 \\ 2006le & -8.9 & 12209(524) & 9(1) & 18176(709) & 7(1) & 23003(855) & 4(1) & ~ & 12882(576) & 68(5) & 20115(806) & 25(4) & ~ & 11159(325) & 12(2) & 24000(862) & 151(13) & ~ & 11005(342) & 0.89(05) & A;B;11 \\
2007F$^*$ & -9.7 & 11868(518) & 6(1) & 17995(702) & 2(1) & 22964(843) & 2(1) & ~ & 12055(348) & 71(5) & 20372(513) & 10(2) & ~ & 10617(296) & 41(4) & 24000(861) & 106(9) & ~ & 11254(324) & 1.03(01) & A;B;2 \\ 2007af & -11.3 & 11183(496) & 35(3) & 17000(674) & 19(2) & 22091(829) & 1(1) & ~ & 12629(632) & 113(12) & 18300(915) & 1(1) & ~ & 11647(318) & 93(10) & 20725(454) & 151(9) & ~ & 11007(315) & 1.20(05) & A;B;C;5 \\
2007bd & -8.1 & 11580(508) & 39(4) & 17500(689) & 20(2) & 24000(887) & 1(1) & ~ & 13709(402) & 130(6) & 18300(775) & 1(1) & ~ & 12459(324) & 93(10) & 21064(457) & 71(4) & ~ & 12557(366) & 1.24(01) & A;B;C;9 \\ 2007bm & -8.2 & 11246(507) & 36(2) & 17170(676) & 12(1) & 21473(813) & 5(1) & ~ & 11145(334) & 89(4) & 18300(774) & 1(1) & ~ & 11064(301) & 85(7) & 20297(444) & 41(2) & ~ & 10500(525) & 1.19(02) & A;B;C;9 \\
2007ca & -10.4 & 11000(491) & 17(2) & 17811(700) & 8(1) & 23102(859) & 1(1) & ~ & 12053(362) & 67(4) & 18564(786) & 10(1) & ~ & \nodata & \nodata & \nodata & \nodata & ~ & 11230(314) & 1.10(01) & A;B;C;9 \\ 2007le & -9.7 & 14623(603) & 9(1) & 20142(776) & 4(1) & 24000(887) & 2(1) & ~ & 13795(637) & 99(8) & 21000(858) & 54(4) & ~ & 14500(349) & 55(6) & 26710(544) & 288(29) & ~ & 12410(370) & 1.01(05) & A;B;C;1 \\
2008ar & -9.1 & 12207(529) & 13(1) & 17145(674) & 7(1) & 21673(819) & 3(1) & ~ & 12187(336) & 93(4) & 19000(492) & 37(5) & ~ & 13247(312) & 58(5) & 23921(681) & 185(13) & ~ & 10634(316) & 1.08(06) & A;B;C;8 \\ 2008bc & -9.7 & 14337(606) & 15(2) & 19000(736) & 6(1) & 23364(865) & 5(1) & ~ & 12933(612) & 90(6) & 19609(815) & 38(3) & ~ & 11774(327) & 41(3) & 24000(862) & 253(11) & ~ & 11600(580) & 0.93(01) & C;9 \\
2008bf & -9.5 & 12000(522) & 14(2) & 18677(723) & 5(1) & 24010(886) & 2(1) & ~ & 12153(360) & 61(5) & 19023(512) & 20(1) & ~ & 12412(319) & 50(5) & 22942(486) & 101(10) & ~ & 11466(336) & 1.03(07) & A;B;C;10 \\ 2008hv & -11.3 & 14314(375) & 36(3) & 18500(721) & 9(1) & 22505(503) & 2(1) & ~ & 13735(424) & 105(6) & 19241(518) & 41(3) & ~ & 14062(347) & 95(9) & 24000(863) & 257(27) & ~ & 10926(312) & 1.25(01) & B;C;11 \\
2009aa & -8.7 & 10381(481) & 28(2) & 16824(683) & 13(1) & 22255(829) & 1(1) & ~ & 10513(316) & 80(4) & 18300(774) & 1(1) & ~ & 10516(294) & 85(7) & 20889(454) & 23(2) & ~ & 10000(500) & 1.12(05) & C;11 \\ 2009ab & -10.8 & 11303(335) & 30(2) & 17730(428) & 16(2) & 23107(512) & 5(1) & ~ & 10921(547) & 73(8) & 18300(915) & 1(1) & ~ & \nodata & \nodata & \nodata & \nodata & ~ & 10839(306) & 1.21(10) & C;3 \\
2009ig & -9.5 & 15000(614) & 8(1) & 18284(440) & 4(1) & 22689(506) & 6(1) & ~ & 14896(668) & 84(6) & 22700(906) & 12(1) & ~ & \nodata & \nodata & \nodata & \nodata & ~ & 13400(670) & 0.89(02) & B;11 \\ 2011by & -12.4 & 11418(333) & 30(2) & 18592(443) & 10(1) & 23001(511) & 1(1) & ~ & 12209(368) & 86(4) & 18300(774) & 7(1) & ~ & 12000(497) & 116(12) & 20650(450) & 79(5) & ~ & 10300(515) & 1.14(03) & D;11 \\
2011df$^+$ & -9.0 & 13711(574) & 9(1) & 18615(722) & 6(1) & 21961(825) & 3(1) & ~ & 12280(366) & 96(5) & 19500(811) & 29(2) & ~ & \nodata & \nodata & \nodata & \nodata & ~ & 11005(348) & \nodata & D; \\ 2011fe$^+$ & -10.0 & 10819(485) & 35(3) & 17522(689) & 14(1) & 22379(837) & 3(1) & ~ & 11917(569) & 99(9) & 20300(835) & 3(1) & ~ & 11537(296) & 108(9) & 19939(431) & 64(4) & ~ & 10444(529) & 1.18(03) & D;12 \\
2012cg$^+$ & -10.0 & 11266(338) & 16(2) & 17726(436) & 8(1) & 23564(517) & 2(1) & ~ & 12278(587) & 64(5) & 18936(790) & 31(3) & ~ & 13052(334) & 78(7) & 21722(466) & 98(5) & ~ & 10000(500) & 0.83(05) & D;11 \\ 2012et$^+$ & -8.0 & 13550(525) & 27(3) & 18500(720) & 11(2) & 23043(948) & 9(1) & ~ & 14051(645) & 119(9) & 18421(779) & 14(1) & ~ & \nodata & \nodata & \nodata & \nodata & ~ & 13053(408) & 1.24(04) & D;11 \\
2012fr$^+$ & -11.5 & 14208(645) & 1(1) & 19312(747) & 3(1) & 26000(948) & 2(1) & ~ & 13400(576) & 40(1) & 22507(857) & 109(10) & ~ & 13000(528) & 9(1) & 27000(547) & 150(14) & ~ & 12200(611) & 0.80(01) & D;2 \\ 2013dy$^*$$^+$ & -10.0 & 11179(331) & 16(2) & 17160(423) & 8(1) & 23184(515) & 6(1) & ~ & 11404(562) & 49(4) & 18417(777) & 5(1) & ~ & 10644(435) & 58(6) & 21985(464) & 52(4) & ~ & 10300(515) & 0.92(01) & D;11 \\
2013gs$^+$ & -9.7 & 13748(573) & 10(1) & 19451(741) & 5(1) & 25227(928) & 1(1) & ~ & 12873(608) & 67(5) & 17561(752) & 56(4) & ~ & 9756(405) & 61(7) & 21112(453) & 185(20) & ~ & 12200(610) & 1.05(01) & D;11 \\ 2013gy & -9.1 & 12106(538) & 32(4) & 16553(710) & 17(2) & 21000(911) & 3(1) & ~ & 13211(392) & 121(6) & 18300(775) & 1(1) & ~ & 12500(512) & 158(16) & 20423(447) & 121(7) & ~ & 11000(316) & 1.20(01) & D;11 \\
2014J$^+$ & -10.5 & 14438(607) & 19(2) & 19972(775) & 7(1) & 23500(872) & 3(1) & ~ & 13920(632) & 114(9) & 20322(816) & 25(2) & ~ & 13043(439) & 52(5) & 24651(693) & 190(19) & ~ & 12050(360) & 1.02(01) & D;11 \\ \enddata
\tablecomments{Spectral parameters measured for O~I $\lambda$7773, Si~II $\lambda$6355, and Ca~II~NIR triplet in the t$\sim$$-$10$\pm$2.5 days spectra of SNe Ia. Note that the velocity, pEW, photospheric component, HVF-I, and HVF-II are abbreviated to V, W, P, H, and S, respectively. The "*" and "$\star$" after the SN name denote the 91T-like and 99aa-like SN Ia, respectively, while the "+" marks the SNe whose spectra used for the measurements of Ca~II~NIR triplet and Si~II $\lambda$6355 (or O~I $\lambda$7773) are taken at slightly different epoches.}
\tablecomments{1-$\sigma$ uncertainties shown in the brackets are in units of 1 km s$^{-1}$ for velocity, 1\AA\ for equivalent width, and 0.01 mag for $\Delta$m$_{15}$(B), respectively.}
\tablecomments{References: A=CfA supernova program \citep{mat08,bl12};  B=Berkeley Supernova Program \citep{si12a}; C=Carnegie Supernova Project \citep[CSP][]{fola13};  D=Tsinghua Supernova Program;
 1= \citep{ha06}; 2=\citep{lira98}; 3=\citep{pa96}; 4=\citep{jh06}; 5= \citep{ga10}; 6=\citep{str02}; 7=\citep{kri03}; 8=\citep{matt05}; 9= \citep{hi09}; 10=\citep{con10}; 11=\citep{wa09a}; 12=\citep{wa08}; 13=\citep{str11}; 14=\citep{hi12}; 15=\citep{mar13}; 16= \citep{fole12}; 17 =\citep{gra15}; 18=\citep{zhao15}; 19=\citep{zh16}; 20=\citep{si12b}; 21=\citep{zh14}; 22=\citep{zh10}}
\tablenotetext{a}{Days since $B-$band maximum light;}
\tablenotetext{b}{Velocity of O~I $\lambda$7773, for the PHO component;} \tablenotetext{c}{Pseudo-equivalent width of O~I $\lambda$7773, for the PHO component;}
\tablenotetext{d}{Velocity of O~I $\lambda$7773, for HVF-I component;} \tablenotetext{e}{Pseudo-equivalent width of O~I $\lambda$7773, for HVF-I component;}
\tablenotetext{f}{Velocity of O~I $\lambda$7773, for the HVF-II component;} \tablenotetext{g}{Pseudo-equivalent width of O~I $\lambda$7773, for the HVF-II component.}
\tablenotetext{h}{Velocity of Si~II $\lambda$6355, for the PHO component.} \tablenotetext{i}{Pseudo-equivalent width of Si~II $\lambda$6355, for the PHO component.}
\tablenotetext{j}{Velocity of Si~II $\lambda$6355, for the HVF component.} \tablenotetext{k}{Pseudo-equivalent width of Si~II $\lambda$6355, for the HVF component.}
\tablenotetext{l}{Velocity of Ca~II~IR triplet, for the PHO component.} \tablenotetext{m}{Pseudo-equivalent width of Ca~II~IR triplet, for the PHO component.}
\tablenotetext{n}{Velocity of Ca~II~IR triplet, for the HVF component.} \tablenotetext{o}{Pseudo-equivalent width of Ca~II~IR triplet, for the HVF component.}
\tablenotetext{p}{Velocity of Si~II $\lambda$6355 measured near the maximum light.} \tablenotetext{q}{Light-curve parameter, B-band light-curve decline rate $\Delta$m$_{15}$(B).}
\tablenotetext{r}{References for both spectra and photometry.} \end{deluxetable}
\newpage
\begin{deluxetable}{lccccccccccccccccccc}
\tabletypesize{\scriptsize}\setlength{\tabcolsep}{0.01in} 
\tablecaption{Fit Results of O I $\lambda$ 7773 \label{T_oxygen}} \tablewidth{0pt}
\tablehead{
\colhead{SN} &\colhead{t\tablenotemark{a}} &\colhead{$V_{P}$\tablenotemark{b}}
&\colhead{$W_{P}$\tablenotemark{c}} &\colhead{$V_{H}$\tablenotemark{e}}
&\colhead{$W_{H}$\tablenotemark{e}} &\colhead{$V_{S}$\tablenotemark{f}}
&\colhead{$W_{S}$\tablenotemark{g}} &\colhead{$R_{b}$\tablenotemark{h}}
\\ \colhead{} &\colhead{d} &\colhead{km~s$^{-1}$} &\colhead{\AA}
&\colhead{km~s$^{-1}$} &\colhead{\AA} &\colhead{km~s$^{-1}$} &\colhead{\AA} &\colhead{~~} } 
\startdata 1990N & -13.9 & 12851(353) & 11(2) & 17946(419) & 12(1) & 21349(795) & 10(2) & 0 \\
1991M & -7.8 & 12501(562) & 40(4) & 17000(675) & 27(2) & 22969(841) & 3(1) & 0.13 \\ 1991T & -10.5 & 11362(331) & 4(1) & 16000(643) & 1(1) & 22539(504) & 2(1) & 0.06 \\
1994D & -11.1 & 13119(361) & 15(2) & 18625(444) & 10(1) & 21719(486) & 2(1) & 0 \\ 1994D & -9.5 & 12601(542) & 22(2) & 17631(697) & 10(1) & 21679(822) & 5(1) & 0 \\
1994D & -8.5 & 12001(507) & 26(2) & 17785(675) & 10(1) & 22284(816) & 1(1) & 0 \\ 1997bq & -11.8 & 13390(302) & 34(3) & 18197(412) & 28(4) & 22603(491) & 14(2) & 0.14 \\
1998dk & -7.1 & 14496(387) & 23(3) & 20364(479) & 15(2) & 23790(527) & 3(1) & 0.07 \\
1998dm & -10.9 & 11702(338) & 21(2) & 17709(429) & 8(1) & 22237(501) & 5(1) & 0.08 \\
1998ef & -8.9 & 11222(509) & 29(4) & 16333(664) & 50(5) & 21660(831) & 14(1) & 0 \\ 1999aa & -11.4 & 12216(539) & 11(1) & 19535(750) & 6(1) & 24103(889) & 4(1) & 0 \\ 1999dk & -7.7 & 13531(460) & 32(2) & 18664(651) & 22(4) & 23000(856) & 1(1) & 0.05 \\ 1999dq & -9.6 & 12355(531) & 5(1) & 18051(699) & 5(1) & 23586(866) & 5(1) & 0 \\ 1999ee & -9.4 & 13000(383) & 5(1) & 18000(706) & 6(1) & 23088(509) & 5(1) & 0.08 \\
1999ee & -7.4 & 13000(378) & 8(1) & 18000(705) & 7(1) & 23025(510) & 6(1) & 0.12 \\
2000dx & -9 & 12501(349) & 18(1) & 17715(431) & 11(1) & 22738(507) & 7(1) & 0 \\ 2000fa & -9.1 & 12011(524) & 20(2) & 20342(779) & 10(1) & 26000(948) & 1(1) & 0 \\
2001el & -9 & 11701(347) & 12(1) & 17847(437) & 8(1) & 22760(506) & 4(1) & 0.09 \\
2002bo & -12.7 & 14626(603) & 31(3) & 20507(782) & 26(2) & 21000(795) & 1(1) & 0 \\
2002cr & -7.6 & 9371(439) & 39(3) & 16117(644) & 16(2) & 20926(792) & 2(1) & 0 \\
2002cs & -9 & 14465(383) & 37(4) & 19000(736) & 24(2) & 23782(523) & 11(2) & 0.02 \\ 2002dj & -12.6 & 16000(645) & 19(2) & 20693(787) & 11(2) & 24000(887) & 1(1) & 0 \\
2002dj & -11.6 & 15487(632) & 43(4) & 20799(797) & 15(2) & 24000(887) & 1(1) & 0 \\
2002dj & -8.6 & 14521(602) & 46(4) & 20000(771) & 15(2) & 24000(887) & 1(1) & 0.06 \\
2002dj & -7.6 & 14000(375) & 35(4) & 20258(782) & 6(1) & 24000(887) & 2(1) & 0.1 \\
2002er & -8.9 & 12962(355) & 35(4) & 19454(465) & 15(2) & 23047(512) & 20(2) & 0.15 \\
2002er & -7.9 & 13183(363) & 36(3) & 20520(596) & 15(2) & 24000(887) & 8(1) & 0.09 \\
2003U & -8.8 & 13054(555) & 51(7) & 17403(738) & 21(2) & 24105(891) & 1(1) & 0 \\
2003cg & -8.4 & 11559(342) & 21(2) & 16576(420) & 12(1) & 21848(494) & 5(1) & 0.11 \\
2003cg & -7.4 & 11763(515) & 22(2) & 16100(647) & 8(1) & 22572(843) & 1(1) & 0.08 \\
2003du & -12.1 & 11762(336) & 16(2) & 17250(424) & 10(1) & 22363(502) & 6(1) & 0.12 \\
2003ek & -10 & 12483(536) & 38(4) & 20853(794) & 23(3) & 26000(948) & 22(2) & 0 \\
2003fa & -9.2 & 10938(488) & 11(1) & 18500(719) & 4(1) & 22412(838) & 4(1) & 0 \\
2003fa & -8.8 & 11009(489) & 9(1) & 18500(719) & 4(1) & 22230(833) & 3(1) & 0 \\
2003kc & -11.4 & 13563(573) & 37(4) & 18611(725) & 31(3) & 21409(807) & 3(1) & 0 \\
2003kf & -9.5 & 12240(535) & 15(1) & 17860(710) & 9(1) & 23000(856) & 5(1) & 0.09 \\
2003kf & -8.8 & 12222(534) & 16(2) & 17651(702) & 9(1) & 23000(856) & 4(1) & 0.08 \\
2004dt & -10.8 & 12157(335) & 43(2) & 17362(418) & 64(7) & 21983(494) & 28(3) & 0.04 \\
2004dt & -7.8 & 10325(318) & 35(4) & 16082(407) & 42(4) & 21199(484) & 20(2) & 0 \\
2004eo & -11.4 & 12500(537) & 21(3) & 17500(690) & 11(2) & 24479(948) & 1(1) & 0.3 \\
2004ey & -8.2 & 12114(523) & 18(2) & 17930(698) & 13(1) & 23000(856) & 2(1) & 0.05 \\
2005cf & -12.7 & 11295(513) & 7(1) & 17213(679) & 8(1) & 23592(873) & 1(1) & 0.12 \\
2005cf & -11.7 & 11186(331) & 12(1) & 16824(416) & 9(1) & 22034(494) & 4(1) & 0.06 \\
2005cf & -11.3 & 11281(332) & 10(1) & 17002(418) & 8(1) & 21877(492) & 4(1) & 0.05 \\
2005cf & -10.7 & 11029(498) & 8(1) & 17242(681) & 5(1) & 23016(857) & 2(1) & 0.08 \\
2005cf & -9.7 & 10968(329) & 8(1) & 16782(416) & 6(1) & 22383(499) & 3(1) & 0.08 \\
2005cf & -8.7 & 10774(485) & 10(1) & 16315(658) & 8(1) & 22081(831) & 5(1) & 0.1 \\
2005cf & -7.7 & 10334(322) & 14(2) & 16334(410) & 8(1) & 21827(489) & 7(1) & 0.12 \\
2005cg & -9 & 11555(509) & 15(2) & 18000(704) & 5(1) & 22210(831) & 2(1) & 0 \\
2005el & -8.1 & 11000(460) & 31(2) & 17078(416) & 14(2) & 22511(503) & 4(1) & 0.04 \\
2005el & -7.1 & 11000(480) & 31(3) & 17171(672) & 12(1) & 22067(818) & 3(1) & 0.03 \\
2005el & -6.9 & 11000(491) & 31(3) & 16555(418) & 12(1) & 21000(488) & 4(1) & 0.02 \\
2005eu & -9.1 & 12447(535) & 5(1) & 18782(728) & 2(1) & 23200(862) & 4(1) & 0 \\
2006X & -11.1 & 15500(660) & 5(1) & 21000(798) & 24(3) & 24040(527) & 7(1) & 0 \\
2006X & -6.8 & 15579(398) & 14(1) & 20536(474) & 24(2) & 24008(856) & 1(1) & 0 \\
2006ax & -10.8 & 11500(506) & 22(2) & 16200(650) & 9(2) & 21410(812) & 2(1) & 0.03 \\
2006ax & -10.4 & 10622(474) & 25(3) & 17102(674) & 11(1) & 22864(851) & 2(1) & 0 \\
2006ax & -8.8 & 10493(466) & 16(2) & 16630(650) & 12(2) & 21959(795) & 2(1) & 0 \\
2006dd & -12.4 & 11156(341) & 21(2) & 17280(427) & 12(2) & 22309(502) & 6(1) & 0.22 \\
2006dm & -7.9 & 10000(460) & 55(6) & 16351(653) & 23(3) & 21000(795) & 7(1) & 0 \\
2006dy & -11.8 & 10556(491) & 38(4) & 18317(722) & 8(1) & 22949(842) & 2(1) & 0.08 \\
2006gr & -9.2 & 13100(556) & 8(1) & 19269(744) & 7(1) & 23148(860) & 3(1) & 0 \\
2006kf & -8.3 & 11991(460) & 64(4) & 17418(673) & 21(3) & 22152(830) & 6(1) & 0.04 \\
2006le & -8.9 & 12209(524) & 9(1) & 18176(709) & 7(1) & 23003(855) & 4(1) & 0 \\
2007F & -9.7 & 11868(518) & 6(1) & 17995(702) & 2(1) & 22964(843) & 2(1) & 0 \\
2007af & -11.3 & 11183(496) & 35(3) & 17000(674) & 19(2) & 22091(829) & 1(1) & 0.11 \\
2007bd & -8.1 & 11580(508) & 39(4) & 17500(689) & 20(2) & 24000(887) & 1(1) & 0 \\
2007bm & -8.2 & 11246(507) & 36(2) & 17170(676) & 12(1) & 21473(813) & 5(1) & 0.06 \\
2007ca & -10.4 & 11000(491) & 17(2) & 17811(700) & 8(1) & 23102(859) & 1(1) & 0 \\
2007ci & -7.1 & 11614(509) & 50(5) & 16000(643) & 20(2) & 22693(845) & 6(1) & 0 \\
2007le & -10.4 & 14623(603) & 6(1) & 20000(766) & 2(1) & 24000(887) & 2(1) & 0 \\
2007le & -9.7 & 14859(616) & 9(1) & 20142(776) & 4(1) & 24000(887) & 2(1) & 0 \\
2007le & -9.5 & 14618(602) & 15(2) & 21000(797) & 4(1) & 24000(887) & 5(1) & 0 \\
2008ar & -9.2 & 12414(535) & 11(1) & 17264(682) & 10(1) & 21827(820) & 3(1) & 0 \\
2008ar & -9.1 & 12000(522) & 14(1) & 17026(666) & 4(1) & 21518(818) & 3(1) & 0 \\
2008bc & -9.7 & 14337(606) & 15(2) & 19000(736) & 6(1) & 23364(865) & 5(1) & 0.09 \\
2008bf & -9.5 & 12000(522) & 14(2) & 18677(723) & 5(1) & 24010(886) & 2(1) & 0 \\
2008hv & -11.3 & 14314(375) & 36(3) & 18500(721) & 9(1) & 22505(503) & 2(1) & 0.2 \\
2009aa & -8.7 & 10381(481) & 28(2) & 16824(683) & 13(1) & 22255(829) & 1(1) & 0 \\
2009aa & -7.7 & 9588(450) & 29(3) & 16000(643) & 16(2) & 21621(814) & 2(1) & 0 \\
2009ab & -10.8 & 11303(335) & 30(2) & 17730(428) & 16(2) & 23107(512) & 5(1) & 0.21 \\
2009dc & -7 & 9000(430) & 47(5) & 16984(673) & 14(2) & 21000(795) & 2(1) & 0 \\
2009ig & -14.4 & 15519(630) & 3(1) & 21500(813) & 1(1) & 24145(891) & 1(1) & 0 \\
2009ig & -12.5 & 15500(630) & 2(1) & 21000(797) & 1(1) & 24192(891) & 1(1) & 0 \\
2009ig & -11.5 & 15682(635) & 2(1) & 20728(694) & 2(1) & 24000(887) & 2(1) & 0 \\
2009ig & -9.5 & 15000(614) & 8(1) & 18284(440) & 4(1) & 22689(506) & 6(1) & 0 \\
2009ig & -8.5 & 14845(387) & 8(1) & 19000(736) & 2(1) & 24000(887) & 3(1) & 0 \\
2011by & -12.4 & 11418(333) & 30(2) & 18592(443) & 10(1) & 23001(511) & 1(1) & 0.04 \\
2011by & -7.3 & 11344(331) & 18(1) & 17637(430) & 12(1) & 22110(498) & 7(1) & 0.03 \\
2011df & -9 & 13711(574) & 9(1) & 18615(722) & 6(1) & 21961(825) & 3(1) & 0 \\
2011fe & -16 & 14347(603) & 40(3) & 19310(749) & 16(1) & 24242(854) & 1(1) & 0 \\
2011fe & -15 & 13439(363) & 28(4) & 18542(448) & 10(1) & 23389(518) & 3(1) & 0 \\
2011fe & -14 & 12367(350) & 33(3) & 18500(443) & 15(1) & 23361(514) & 2(1) & 0 \\
2011fe & -13 & 12015(351) & 34(2) & 18086(437) & 12(1) & 23451(517) & 1(1) & 0 \\
2011fe & -12 & 11936(519) & 33(2) & 18022(706) & 10(1) & 22445(840) & 3(1) & 0 \\
2011fe & -11 & 10849(486) & 39(4) & 17647(691) & 14(2) & 22611(844) & 3(1) & 0 \\
2011fe & -10 & 10800(485) & 35(3) & 17522(689) & 13(1) & 22379(837) & 2(1) & 0 \\
2011fe & -9 & 10819(323) & 27(2) & 17183(421) & 14(1) & 21868(492) & 6(1) & 0 \\
2011fe & -8 & 11163(331) & 26(1) & 17082(420) & 10(1) & 21392(487) & 3(1) & 0 \\
2011fe & -7 & 10797(476) & 26(3) & 16748(678) & 14(1) & 21663(809) & 6(1) & 0.01 \\
2012cg & -13.3 & 12719(544) & 22(2) & 18500(720) & 13(1) & 23144(860) & 1(1) & 0 \\
2012cg & -11.6 & 12737(544) & 19(2) & 17685(695) & 14(1) & 22200(832) & 3(1) & 0 \\
2012cg & -10 & 11266(338) & 16(2) & 17726(436) & 8(1) & 23564(517) & 2(1) & 0 \\
2012cg & -8.6 & 11360(513) & 13(1) & 17739(703) & 9(1) & 23356(868) & 4(1) & 0 \\
2012cg & -8.3 & 11326(332) & 12(1) & 17667(429) & 7(1) & 23038(512) & 4(1) & 0 \\
2012et & -8 & 13550(525) & 27(3) & 18500(720) & 11(2) & 23043(948) & 9(1) & 0 \\
2012fr & -12.5 & 14500(599) & 1(1) & 21000(797) & 3(1) & 23269(864) & 4(1) & 0 \\
2012fr & -11.5 & 14208(645) & 1(1) & 19312(747) & 3(1) & 26000(948) & 2(1) & 0 \\
2012fr & -8.3 & 13672(572) & 2(1) & 18971(737) & 4(1) & 22325(855) & 1(1) & 0 \\
2012fr & -8 & 13000(552) & 10(1) & 17294(425) & 5(1) & 21000(795) & 2(1) & 0 \\
2012fr & -7.7 & 13116(558) & 8(1) & 18769(729) & 3(1) & 23500(871) & 2(1) & 0 \\
2012fr & -7.5 & 13406(364) & 6(1) & 19000(736) & 5(1) & 23500(871) & 3(1) & 0 \\
2012fr & -6.8 & 12766(567) & 14(1) & 18601(730) & 7(1) & 25802(944) & 1(1) & 0.03 \\
2012fr & -6.7 & 12783(563) & 13(1) & 18740(724) & 9(1) & 24251(891) & 3(1) & 0 \\
2013ah & -7 & 10083(304) & 39(4) & 19757(462) & 18(2) & 26000(948) & 11(1) & 0 \\
2013dy & -15.8 & 13732(590) & 10(1) & 18000(705) & 7(1) & 21752(806) & 1(1) & 0 \\
2013dy & -12 & 11147(347) & 15(1) & 17500(427) & 6(1) & 23923(526) & 2(1) & 0.1 \\
2013dy & -10.5 & 11375(333) & 15(2) & 17446(427) & 8(1) & 23378(517) & 6(1) & 0.06 \\
2013dy & -9.5 & 10983(328) & 17(2) & 16873(419) & 7(1) & 22990(512) & 5(1) & 0.09 \\
2013dy & -7.5 & 10781(333) & 20(2) & 16400(418) & 9(1) & 21617(495) & 5(1) & 0.15 \\
2013gs & -9.7 & 13748(573) & 10(1) & 19451(741) & 5(1) & 25227(928) & 1(1) & 0 \\
2013gs & -8.4 & 13500(568) & 9(1) & 18638(439) & 8(1) & 25356(546) & 2(1) & 0.19 \\
2013gs & -7.5 & 13200(559) & 21(2) & 19742(759) & 9(1) & 26000(948) & 2(1) & 0.03 \\
2013gy & -9.1 & 12106(538) & 32(4) & 16553(710) & 17(2) & 21000(911) & 3(1) & 0.12 \\
2014J & -11 & 14952(614) & 19(2) & 21000(797) & 7(1) & 24000(887) & 2(1) & 0 \\
2014J & -10 & 13923(599) & 18(2) & 18943(752) & 7(1) & 23000(857) & 4(1) & 0 \\
2014J & -7.7 & 13459(379) & 23(2) & 19514(480) & 9(1) & 22782(510) & 1(1) & 0 \\
\enddata

\tablecomments{Spectral parameters measured for the absorption O~I $\lambda$7773 in the early spectra of SNe Ia . The abbreviations are similar to those used in Table 1, with V, W, P, H, and S representing the velocity, pEW, photospheric component, HVF-I, and HVF-II, respectively. 1-$\sigma$ uncertainties shown in the bracket are in unit of 1 km s$^{-1}$ and 1 \AA\ for velocity and equivalent width, respectively.}
\tablenotetext{a}{Days since the $B-$band maximum light;}
\tablenotetext{b}{Velocity of O~I $\lambda$7773, for the PHO component;}
\tablenotetext{c}{Pseudo-equivalent width of O~I $\lambda$7773, for the PHO component;}
\tablenotetext{d}{Velocity of O~I $\lambda$7773, for HVF-I component;}
\tablenotetext{e}{Pseudo-equivalent width of O~I $\lambda$7773, for HVF-I component;}
\tablenotetext{f}{Velocity of O~I $\lambda$7773, for the HVF-II component;}
\tablenotetext{g}{Pseudo-equivalent width of O~I $\lambda$7773, for the HVF-II component.}
\tablenotetext{h}{The blending ratio between the telluric lines and the photospheric component of O~I $\lambda$7773, defined as pEW$^{telluric}_{blending}$/pEW$_{PHO}^{O}$, where pEW$^{telluric}_{blending}$ is the pEW of the telluric absorptions that is overlapped with the photospheric O~I $\lambda$7773, and pEW$_{PHO}^{O}$ refers to the pEW of photospheric component of O I $\lambda$7773.}
\end{deluxetable}

\newpage

\begin{deluxetable}{lcccclcccclcccccccc}
\tabletypesize{\scriptsize}\setlength{\tabcolsep}{0.05in}
\tablecaption{Fit Results of C~II $\lambda$6580 \label{T_carbon}}
\tablewidth{0pt} \tablehead{ 
\colhead{SN} &\colhead{t} &\colhead{V} &\colhead{W} &\colhead{}
&\colhead{SN} &\colhead{t} &\colhead{V} &\colhead{W} &\colhead{}
&\colhead{SN} &\colhead{t} &\colhead{V} &\colhead{W}
\\
\colhead{}  &\colhead{d} &\colhead{km~s$^{-1}$} &\colhead{\AA}
&\colhead{~~} &\colhead{}  &\colhead{d} &\colhead{km~s$^{-1}$}
&\colhead{\AA} &\colhead{~~} &\colhead{}  &\colhead{d}
&\colhead{km~s$^{-1}$} &\colhead{\AA} }

\startdata
1990N & -13.9 & 13496(675) & 8.2(0.9) & ~ & 2005cf & -12.5 & 12048(362) & 1.4(1) & ~ & 2008Z & -9.4 & 14000(700) & 7.1(0.8) \\
1994D & -9.5 & 13114(396) & 3.9(1) & ~ & 2005cf & -11.7 & 11886(362) & 1.5(1) & ~ & 2008bf & -9.5 & 12917(388) & 2.7(1) \\
1994D & -8.5 & 13060(382) & 4.6(1) & ~ & 2005el & -8.1 & 12130(364) & 4.2(1) & ~ & 2009F & -5.8 & 12145(608) & 6.2(0.7) \\
1998aq & -9.9 & 12072(604) & 3.7(0.4) & ~ & 2005el & -7.1 & 11791(352) & 4(1) & ~ & 2009dc & -7 & 10351(314) & 5.7(1) \\
1998dm & -10.9 & 12580(629) & 1.7(0.2) & ~ & 2005eu & -9.7 & 12487(625) & 2.6(0.3) & ~ & 2009ig & -15 & 9304(466) & 0.5(0.1) \\
1999by & -6.2 & 10454(523) & 1.3(0.2) & ~ & 2005iq & -6.4 & 11889(595) & 3.2(0.4) & ~ & 2011by & -10 & 12821(390) & 3.1(1) \\
1999cp & -12 & 12688(635) & 6.3(0.7) & ~ & 2006D & -6.3 & 12010(601) & 6.8(0.7) & ~ & 2011fe & -13 & 11732(360) & 3.2(1) \\
2002cr & -11.3 & 11007(330) & 2.7(1) & ~ & 2006ax & -11.1 & 11326(567) & 5.2(0.6) & ~ & 2011fe & -12 & 11778(354) & 2.1(1) \\
2002cr & -7.6 & 10563(316) & 2.4(1) & ~ & 2006dd & -12.4 & 12436(368) & 2.9(1) & ~ & 2011fe & -11 & 11402(346) & 1.9(1) \\
2003du & -14.1 & 12722(382) & 2.8(1) & ~ & 2006dy & -11.8 & 11914(362) & 3.1(1) & ~ & 2011fe & -10 & 11285(332) & 3.6(1) \\
2003du & -12.8 & 12392(372) & 3.9(1) & ~ & 2006gz & -9.6 & 13622(682) & 14.6(1.5) & ~ & 2011fe & -9 & 11094(334) & 2.1(1) \\
2003du & -12.1 & 12311(370) & 1.9(1) & ~ & 2006le & -7.9 & 12284(615) & 3.7(0.4) & ~ & 2011fe & -7 & 10792(324) & 1.8(1) \\
2003du & -10.8 & 12221(366) & 2(1) & ~ & 2007F & -9.7 & 12640(380) & 1.2(1) & ~ & 2012cg & -10 & 12668(382) & 4.9(1) \\
2003du & -9.9 & 12009(360) & 2.6(1) & ~ & 2007af & -11.3 & 12031(366) & 1.2(1) & ~ & 2012cg & -8.6 & 12400(595) & 3.1(1) \\
2003du & -8.9 & 12378(372) & 2(1) & ~ & 2007bm & -9.3 & 11436(344) & 2.3(1) & ~ & 2012cu & -7 & 12150(608) & 8.1(0.9) \\
2003kf & -9.4 & 12878(644) & 2.4(0.3) & ~ & 2007bm & -8.5 & 11260(342) & 1.6(1) & ~ & 2013dy & -15.8 & $>$13000 & $>$20 \\
2004bw & -10.5 & 12301(616) & 10.5(1.1) & ~ & 2007le & -10.7 & 13607(681) & 0.6(0.1) & ~ & \nodata & \nodata & \nodata & \nodata \\
2005cf & -12.7 & 12048(362) & 1.4(1) & ~ & 2008Q & -7.3 & 14000(700) & 6.5(0.7) & ~ & \nodata & \nodata & \nodata & \nodata \\
\enddata
\tablecomments{Spectral parameters measured for C~II $\lambda$6580 in early spectra of SNe Ia. Velocity is abbreviated to V, and pEW is further abbreviated to W due to limited space. 1-$\sigma$ uncertainties shown in brackets are in unit of 1 km s$^{-1}$ and 1 \AA\ for velocity and equivalent width, respectively.}
\end{deluxetable}

\end{document}